\documentclass[showpacs,twocolumn,prb]{revtex4-1}
\usepackage{graphicx}
\graphicspath{{figures/}} 

\newcommand{\be}{\begin{equation}}
\newcommand{\ee}{\end{equation}}
\newcommand{\bea}{\begin{eqnarray}}
\newcommand{\eea}{\end{eqnarray}}
\newcommand{\bwt}{\begin{widetext}}
\newcommand{\ewt}{\end{widetext}}

\usepackage{hyperref}
\hypersetup{
    colorlinks=true,
    linkcolor=blue,
    filecolor=magenta,      
    urlcolor=blue,
}

\usepackage{color} 
\newcommand{\itt}{\it}
\newcommand{\black}{\textcolor{black}}
\newcommand{\red}{\textcolor{black}}

\usepackage{comment}
\def\comment#1{}
\newcommand{\com}{\black}

\begin{document}

\title{A systematic study of the superconducting critical temperature in two and three dimensional tight-binding models\red{: a possible scenario 
for superconducting H$_3$S ?}}

\author{Thiago X. R. Souza}
\affiliation{Department of Physics, University of Alberta, Edmonton, AB, Canada T6G~2E1,\\ {\rm and} \\
Departamento de Fisica, Universidade Federal de Sergipe, 49100-000 Sao Cristovao, SE, Brazil}
\author{F. Marsiglio}
\affiliation{Department of Physics, University of Alberta, Edmonton, AB, Canada T6G~2E1}

\begin{abstract}
Ever since BCS theory was first formulated it was recognized that a large electronic density of states at the Fermi level
was beneficial to enhancing $T_c$. The A15 compounds and the high temperature cuprate materials both have had an
enormous amount of effort devoted to studying the possibility that such peaks play an important role in the high critical temperatures
existing in these compounds. Here we provide a systematic study of the effect of these peaks on the superconducting
transition temperature for a variety of tight-binding models of simple structures, both in two and three dimensions.
In three dimensions large enhancements in $T_c$ can occur, due to van Hove singularities \red{\it that result in divergences in the
density of states}. Furthermore, even in more
realistic structures, where the van Hove singularity disappears, large enhancements in $T_c$ continue due to the presence
of `robust' peaks in the densities of states. \red{Such a peak, recently identified in the bcc structure of H$_3$S, is likely the result of such
a van Hove singularity.} \red{In certain regimes,} anomalies in the isotope coefficient are also expected.
\end{abstract}

\date{\today }
\maketitle

\section{introduction}

The weak coupling Bardeen-Cooper-Schrieffer (BCS) \cite{bardeen57} expression for the superconducting transition temperature $T_c$
is
\be
T_c \sim \omega_D e^{-1/[g(\epsilon_F)V]}
\label{bcs_weak_tc}
\ee
(we set $\hbar = k_B = 1$) where $\omega_D$ is the typical (Debye) phonon frequency, $V$ is the attractive interaction strength, and $g(\epsilon_F)$ is the electron density
of states at the Fermi level. This simple expression makes clear that a high value of the density of states at the Fermi energy is desirable for high
$T_c$, and has served to motivate a directed search for high $T_c$ materials for more than half a century. Some understanding of the impact on $T_c$
has come historically from a study of the A15 compounds, where experiments suggested that various `anomalous' superconducting properties
in these compounds could be explained by peaks (or in some cases valleys) in the electron density of states near the Fermi level.

Indeed, as early as 1967 Labb\'e et al.\cite{labbe67} suggested that sharp peaks in the electronic density of states could explain the high $T_c$ and
low isotope effects in some A15 compounds. They adopted a density of states with a square-root singularity, reminiscent of the result obtained in
one dimension. Since that time, Nettel and Thomas\cite{nettel77} and Horsch and Rietschel\cite{horsch77} developed this model further in the context of Eliashberg theory, again with an eye towards explaining the high critical temperatures of some of the A15 compounds. 
Follow-up work by Lie and Carbotte\cite{lie78}, Ho et al.\cite{ho78}, Pickett\cite{pickett80} and Mitrovi\'c and Carbotte\cite{mitrovic83} served to establish
the importance of peaked structures in the electronic density of states near the Fermi level for the critical temperatures in the A15 superconductors. 

In the mid 1980's the possibility of enhancing the superconducting critical temperature through a two-dimensional structure was advanced by Hirsch and
Scalapino,\cite{hirsch86} and these authors also used Monte Carlo simulations and high-order perturbation corrections to support their claims. They found
enhanced superconductivity when the Fermi level was near a singularity, particularly in the weak coupling regime. These ideas were further developed with
the discovery of high temperature superconductivity in 1986, and several papers\cite{labbe87,tsuei90} subsequently explored some of the consequences of
a two-dimensional van Hove singularity for superconductivity. Rather than recount a detailed history of the various calculations, we refer the reader to review papers,
a comprehensive one in 1997,\cite{markiewicz97} and a more recent review\cite{bok12} focussed on the A15 compounds. While the early work focussed
on a square-root singularity, most of the work in the last 30 years has almost exclusively utilized a density of states with a logarithmic divergence, motivated by
the two-dimensional tight-binding model. A notable exception is the extended saddle point singularity pointed out through density functional theory calculations in 1991\cite{andersen91} and observed through ARPES measurements and modelled in 1993.\cite{abrikosov93} The extended saddle point results in a 
one-dimensional-like square-root singularity in the electronic density of states.

The theoretical description of these various scenarios has focussed on the situation where the Fermi energy lies close to the singularity in the density of states. 
In this paper we wish to do two things. First, we will extend these calculations to all electron densities, in the vicinity of the singularity, and well away from it. Our
results will be numerical, and will account self-consistently for changes in the chemical potential as the electron density and coupling strength of the pairing
interaction varies. These calculations will be performed for a two-dimensional tight-binding model on a square lattice,
where a logarithmic singularity in the electronic density of states always exists.

Secondly, we will extend these calculations to three dimensions. Of course van Hove anomalies also exist in three dimensions. Jelitto\cite{jelitto69} showed
long ago that for the body-centred cubic (bcc) and face-centred cubic (fcc) lattice structures these anomalies result in singularities in the density of states as
well. As this important result appears to be under-appreciated, we review some of his results in the Appendix. \red{Finally, we note that the density of states
for the bcc lattice, with a non-negligible next-nearest-neighbor (NNN) hopping amplitude, renders a density of states with a significant and `robust' peak, very similar
to one recently calculated\cite{duan14,bernstein15} with density functional theory for the newly discovered superconductor, H$_3$S.\cite{drozdov15} We find a significant enhancement of $T_c$ for electron densities obtained for the chemical
potential close to the energy of this peak.}
In summary, \red{while the bulk of this paper is devoted to} a comprehensive survey for $T_c$ (and in some cases the isotope coefficient
and the superconducting order parameter), as a function of electron density and coupling strength, in both two and three dimensions, for a variety of ``cubic''
lattice structures, \red{we find that the bcc structure itself results in a substantial enhancement of $T_c$}.

It is probably best to specify the simplifying assumptions that we utilize: (i) We assume a momentum independent pairing interaction, and hence this study is
confined to a superconducting order parameter with s-wave symmetry. (ii) We will adopt a non-retarded framework for the interaction, i.e. we will use the BCS
formalism, rather than the Eliashberg formalism. Many authors (see, e.g. Ho et al.\cite{ho78}) have pointed out that retardation effects will smear the effective
electronic density of states, so that a BCS-like treatment will tend to over-estimate the effects of a singularity in the density of states. This is understood here, and it is desirable to have a follow-up study similar to this one
based on the Eliashberg formalism.\cite{remark0} (iii) We will focus on a metal in which a single band crosses the Fermi level;
furthermore, we will adopt a tight-binding model to describe the dispersion of this band, and correlation effects in the normal state are assumed to be absent.
(iv) While we will adopt analytical approximations from time to time these will be for illustrative purposes only --- all our main results will be numerically exact, with no weak coupling approximations, for example. The one exception is that at the band edges we do not concern ourselves with possible strong coupling effects. These effects will give rise to Bose condensation physics dominating over BCS pairing (i.e. condensation arises not from pairing {\it per se}, but from phase coherence);
however, since the theoretical description of this crossover is not universally agreed upon,\cite{nozieres85,marsiglio15} for present purposes
we simply use the BCS formalism in this very small regime as well.

The outline is as follows. In the next section we provide a concise formulation of the equations we solve, both at the critical temperature $T_c$, and at
temperatures below $T_c$. In the following section we focus on the two dimensional square lattice, first with nearest neighbour hopping only, and then
with next-nearest neighbour hopping. We examine $T_c = T_c(n,V,\omega_D)$, where $n$ is the electron density, $V$ is the coupling strength, and
$\omega_D$ is used as a cutoff, representing the Debye frequency of the phonons. We also examine
the isotope coefficient (to be defined below) and the superconducting order parameter, $\Delta$. For the most part $\Delta(n,V,\omega_D)$ tracks 
$T_c(n,V,\omega_D)$, and the temperature dependence of $\Delta$ is essentially indistinguishable from that achieved with a constant density of states. Results
for a constant density of states have previously been presented\cite{marsiglio92} within the Eliashberg\cite{eliashberg60,marsiglio08} formalism. These results, recalculated with the
much simpler BCS formalism, will provide a baseline for comparisons.

The fourth section will focus on the three dimensional cubic lattices, simple cubic (sc), body-centred cubic (bcc) and face-centred cubic (fcc). 
The first two have particle-hole symmetry, while the third
does not, and the singularity in the electron density of states for the fcc lattice lies at the upper end of the spectrum. We also consider the impact of next
nearest neighbour hopping in all three cases. Somewhat surprisingly, in the bcc and fcc cases, while the singularity is removed, a robust peak
remains, and considerable enhancement of $T_c$ occurs. Equally surprisingly, in the sc case, turning on the next nearest neighbour hopping moves
the density of states towards one with a singularity.

\red{Finally, we point out that for a particular range of NNN hopping amplitude, the density of states resembles that calculated\cite{quan16,sano16} 
with density functional theory, and leads to a significant enhancement of $T_c$.}

\section{The pairing formalism}

The BCS equations are written as\cite{schrieffer64,tinkham96}

\be
\Delta_k = -{1 \over N} \sum_{k^\prime} V_{kk^\prime} { \Delta_{k^\prime} \over 2 E_{k^\prime}} \left[ 1 - 2f(E_{k^\prime}) \right],
\label{bcs1}
\ee
and
\be
n = {1 \over N} \sum_{k^\prime} \left[ 1 - {\epsilon_{k^\prime} - \mu \over E_{k^\prime}} \left( 1 - 2f(E_{k^\prime})\right) \right],
\label{bcs2}
\ee
with
\be
E_k \equiv \sqrt{  (\epsilon_{k^\prime} - \mu)^2 + \Delta^2_{k^\prime}}.
\label{bcs3}
\ee
Here, $\Delta_k$ is the superconducting order parameter; this parameter goes to zero at the critical temperature $T_c$. $N$ is the number of
unit cells in the lattice and the summation over $k$-points is to cover the entire First Brillouin zone (FBZ). In principle this summation also covers
all bands in the FBZ, but as specified in our assumptions we focus on one band only, in which the Fermi energy lies. The pairing interaction, to
be specified further below, is given by $V_{kk^\prime}$. Note that the dependence on the centre-of-mass momentum $q$ is absent, so that
this is the interaction for the so-called ``reduced BCS'' Hamiltonian. We have also adopted the convention that an attractive interaction will be
negative, so that Eq. (\ref{bcs1}) has a minus sign. The chemical potential is denoted by $\mu$; it will generally be altered by the presence of the
superconducting state, although in practice, in weak and intermediate coupling situations it will change only by a very small amount. By using Eq. (\ref{bcs2})
we take these changes into account in order to preserve the electron density, $n$, as the various parameters, such as temperature, or even the
``fixed'' parameters like $\omega_D$, are varied. Finally all the temperature dependence is included through the Fermi-Dirac distribution function,
$f(x) \equiv 1/[{\rm exp}(\beta x) + 1]$, where $\beta \equiv 1/[k_BT]$ is the inverse temperature, with $k_B$ the Boltzmann constant.

In addition we need to specify an energy dispersion, $\epsilon_k$. We adopt the tight binding model, so for example, with nearest-neighbour (NN)
hopping only, we obtain
\bea
\epsilon_k &=& -2t\left[ {\rm cos}(k_xa) + {\rm cos}(k_ya) \right] \phantom{aaaaaaaaaaa} {\rm [2D]} \label{2d} \\
\epsilon_k &=& -2t_s\left[ {\rm cos}(k_xa) + {\rm cos}(k_ya) +{\rm cos}(k_za) \right] \phantom{aa} {\rm [sc]} \label{sc} \\
\epsilon_k &=& -8t_b\left[ {\rm cos}({k_xa \over 2})  {\rm cos}({k_ya \over 2}) {\rm cos}({k_za \over 2}) \right] \phantom{aaaa} {\rm [bcc]} \label{bcc} \\
\epsilon_k &=& -4t_f \biggl[ {\rm cos}({k_xa \over 2}) {\rm cos}({k_ya \over 2}) + {\rm cos}({k_xa \over 2}) {\rm cos}({k_za \over 2})  \nonumber \\
&& \phantom{aaa} + {\rm cos}({k_ya \over 2}) {\rm cos}({k_za \over 2})  \biggr] \phantom{aaaaaaaaaaaa} {\rm [fcc]}
\label{fcc}
\eea
for the four structures considered, where $a$ is the nearest neighbour distance in the 2D and (sc) cases, and is the length of the cube in the bcc and fcc cases, 
containing 8 atoms at each vertex along with one in the centre (bcc) and six on the face centres (fcc). Also, $t, t_s, t_b$, and $t_f$ are the nearest neighbour hopping parameters for the 2D square, 3D simple cubic, 3D bcc, and 3D fcc lattices, respectively. Note that these have bandwidths $W$ of $8t$, $12t_s$, $16t_b$, 
and $16t_f$, respectively. In the main text and figures that follow, we will generally use `{\it t}' to designate the NN hopping, and `$t_2$' to
designate the next-nearest neighbour (NNN) hopping parameter (see Appendix). Thus, unless necessary to distinguish the various cases, we
will drop the additional subscript, $s$, $b$, and $f$, and retain them only as needed. These dispersions are further discussed in the Appendix.

At this point the main simplifying assumption in the ensuing calculations is that the pairing interaction is essentially local, so that the pairing
interaction is independent of momentum. We wish to retain the notion that pairing is via boson exchange, and with the phonon mechanism in
mind following BCS,\cite{bardeen57} we want to include a feature that requires the two electrons to have single particle energies that are no further
than $\hbar \omega_D$ part from one another. This is difficult to implement in practice, so instead we adopt the standard model that restricts each of the single
particle energies to be within $\hbar \omega_D$ of the chemical potential, $\mu$. That is,
\be
V_{kk^\prime} = -V \theta \left[ \hbar \omega_D - |\epsilon_k - \mu | \right] \theta \left[ \hbar \omega_D - |\epsilon_{k^\prime} - \mu | \right] 
\label{vpair}
\ee
where $\theta[x] \equiv 0$ for $x<0$ and $\theta[x] \equiv 1$ for $x> 0$ is the Heaviside step function, and $V>0$ implies that this is an attractive interaction
potential. With this model in place, the order parameter becomes non-zero only for $|\epsilon_k - \mu| < \hbar \omega_D$, and its value is independent of
momentum.\cite{remark1}

Because of the simplicity of this model potential, one can rewrite the momentum sums in Eqs. (\ref{bcs1},\ref{bcs2}) in terms of the electronic density of states,
$g(\epsilon)$ (see the Appendix). Then these equations become
\be
{1 \over V} = \int_{\mu_-}^{\mu_+} \ d\epsilon g(\epsilon) {{\rm tanh}[\beta E(\epsilon)/2] \over 2E(\epsilon)}
\label{bcs1a}
\ee
and
\be
n = \int_{\epsilon_{\rm min}}^{\epsilon_{\rm max}}   \ d\epsilon g(\epsilon) \left[ 1 - {(\epsilon - \mu) \over E(\epsilon)}
{\rm tanh}[\beta E(\epsilon)/2] \right],
\label{bcs2a}
\ee
with $E(\epsilon)= \sqrt{  (\epsilon - \mu)^2 + \Delta^2}$. Here, the integration limits in Eq. (\ref{bcs1a}) are normally $\mu_- = \mu - \hbar \omega_D$
and $\mu_+ = \mu + \hbar \omega_D$, while those in Eq. (\ref{bcs2a}) are $\epsilon_{\rm min}$, the band energy at the bottom of the band,
and $\epsilon_{\rm max}$, the band energy at the top of the band.  An exception occurs when the chemical potential is close to one of the band edges. 
In this case, the integration is cut off by the band edge, so more accurately, $\mu_- \equiv{\rm max}[ \mu - \hbar \omega_D, \epsilon_{\rm min}]$, and 
$\mu_+ \equiv{\rm min}[ \mu + \hbar \omega_D, \epsilon_{\rm max}]$.

Equations (\ref{bcs1a},\ref{bcs2a}) represent two non-linear equations for the unknowns $\Delta$ and $\mu$, given the parameters $V$ and $n$. At
zero temperature the hyperbolic tangent function is replaced by unity; at $T_c$ the order parameter goes to zero so the problem is slightly different. One
then has to find the temperature and the chemical potential at which both these equations are satisfied. These equations are
\be
{1 \over V} = \int_{\mu_-}^{\mu_+} \ d\epsilon g(\epsilon) {{\rm tanh}[\beta_c (\epsilon -\mu)/2] \over 2(\epsilon - \mu)} \phantom{aaaaa} [T=T_c]
\label{bcs1tc}
\ee
and
\be
n = 2\int_{\epsilon_{\rm min}}^{\epsilon_{\rm max}}   \ d\epsilon g(\epsilon) f(\epsilon - \mu),  \phantom{aaaaaaaaaa}  [T=T_c]
\label{bcs2tc}
\ee
where $\beta_c \equiv 1/[k_BT_c]$. Numerical results in subsequent sections are the result of an iterative solution to these equations.

\bigskip

\section{Two Dimensions}

As detailed in the Appendix, the electron density of states for a two-dimensional tight-binding model with nearest neighbour hopping only is
\be
g_{\rm 2D}(\epsilon) = {1 \over 2 \pi^2 t a^2} K\left[ 1 - \left({\epsilon \over 4 t}\right)^2\right],
\label{dense_2d}
\ee
where $K(m) \equiv \int_0^{\pi/2} \ d\theta \ \left({1 - m \ {\rm sin}^2\theta}\right)^{-1/2}$ is the complete elliptic integral of the first kind.\cite{nist10}
This density of states is well approximated by the expression
\be
g_{\rm 2D}(\epsilon) \approx {1 \over 2 \pi^2 t a^2} {\rm log}({16t \over |\epsilon|});
\label{dense_2d_app}
\ee
Eq.~(\ref{dense_2d_app}) is the asymptotic form of Eq.~(\ref{dense_2d}) as 
$\epsilon \rightarrow 0$, and is often used in lieu
of  Eq.~(\ref{dense_2d}). Both are shown in Fig.~\ref{figA1} in the Appendix, along with a numerical evaluation of the density of states when the next-nearest-neighbour
hopping is included as well. As discussed in the Appendix, a simple numerical routine can efficiently and accurately
evaluate complete elliptic integrals, so we proceed with the full form, Eq.~(\ref{dense_2d}).

\subsection{ Numerical results}

\begin{figure}[tp]
\begin{center}
\includegraphics[height=3.7in,width=3.2in,angle=-90]{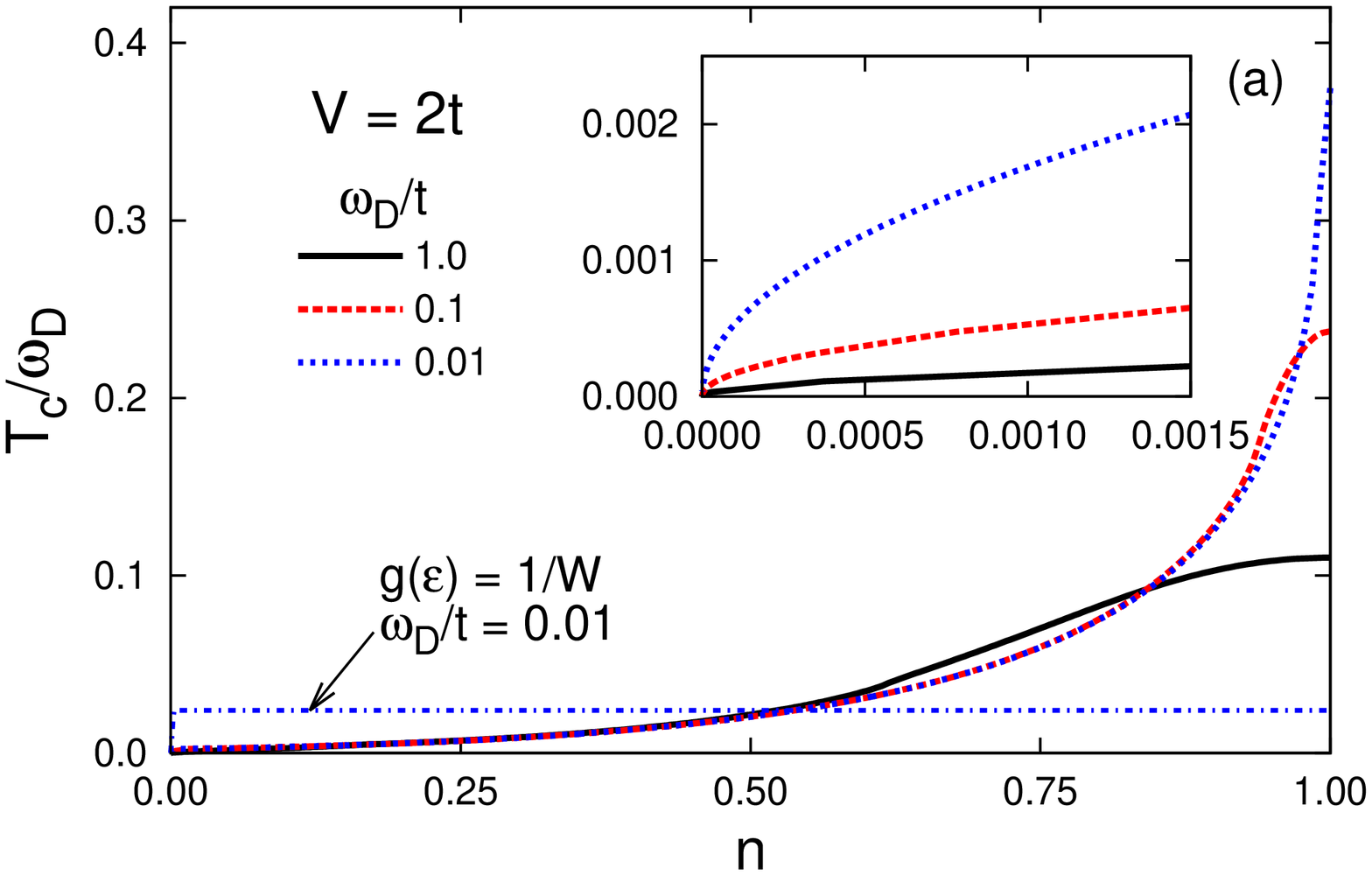}
\includegraphics[height=3.2in,width=2.5in,angle=-90]{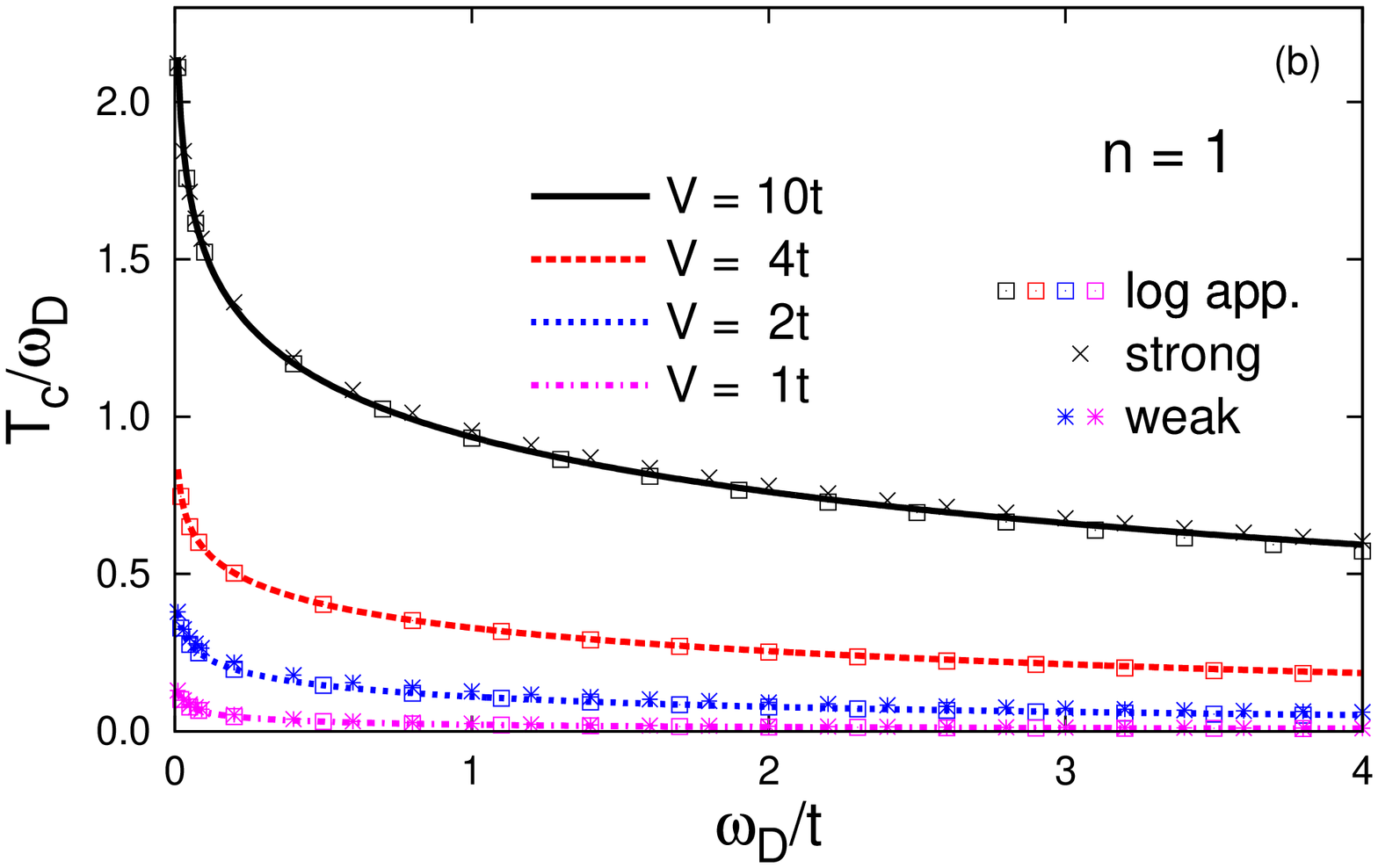}
\end{center}
\caption{(a) Plot of $T_c/\omega_D$ vs. $n$ for $0 < n < 1$ (the results for $2 > n > 1$ are symmetric) for $V/t = 2$, and 
$\omega_D/t = 1.0, 0.1, 0.01$. Also shown is the result for
a constant density of states, $g(\epsilon) = 1/W$, where $W = 8t$ is the electron bandwidth. This result is not sensitive to $\omega_D$ except at the band
edges. In the inset we show an expanded view of the region near zero density (also the case near $n = 2$), showing how $T_c \propto \sqrt{n}$ due to
the lower band edge taking the place of $\mu -\omega_D$ for the lower cutoff. There is a clear enhancement near the van Hove singularity, especially for
small $\omega_D/t$. (b) $T_c/\omega_D$ vs. $\omega_D/t$  for $n=1$ for several values of the coupling strength $V$, along with several
approximations discussed in the text.}
\label{fig1}
\end{figure}

We show in Fig.~(\ref{fig1}a) $T_c/\omega_D$ (we set $\hbar = k_B = 1$) vs. electron density $n$, for a relatively weak coupling situation, $V/t = 2$, for
three different values of $\omega_D/t = 1.0$, $0.1$, and $0.01$. The latter two values are more realistic as in general, $\omega_D << t$, i.e. phonon
energy scales are much smaller than electronic energy scales. The curves provide results for the complete self-consistent solution without approximation,
i.e. using the density of states from Eq.~(\ref{dense_2d}). There is a clear enhancement of $T_c$, especially near the van Hove singularity. and the
enhancement is most pronounced for smaller values of $\omega_D$. Also shown is the result for a constant density of states; for electron densities
near half-filling this shows that $T_c$ can be enhanced by more than an order of magnitude. 
In Fig.~(\ref{fig1}b) we show the same quantity as a function of $\omega_D/t$,
now for a variety of values of $V/t$, all for half-filling (shown with curves). $T_c$ will tend to increase as a function of $\omega_D$, but we have 
plotted the ratio, $T_c/\omega_D$ vs. $\omega_D/t$, which shows an enhancement of $T_c/\omega_D$ as $\omega_D \rightarrow 0$. 

\subsection{Analytical results}

Analytical results are possible through a series of simplifications, as follows. First, we focus on half-filling, $n=1$. This means that the chemical potential
remains fixed at $\mu = 0$, independent of temperature. Second, we adopt the approximation given in Eq.~(\ref{dense_2d_app}), which, based on the
comparisons of the density of states given in the Appendix, we anticipate will be very accurate. Indeed, this is the case, as indicated by the results depicted
with square symbols in Fig.~(\ref{fig1}b). In particular, these results are always very accurate as $\omega_D \rightarrow 0$, as this is where the density of states at the singularity
is most important; this is also where the approximation Eq.~(\ref{dense_2d_app}) is most accurate.

A so-called "strong-coupling" approximation to the BCS equation is obtained as follows. We assume $\omega_D/T_c <<1$, which means that the
hyperbolic tangent function can be linearized. The remaining integral is then elementary, so that
\be
{T_c \over \omega_D} \approx {V \over 4 \pi^2 t}\bigl[{\rm log}({16t \over \omega_D}) + 1 \bigr] \phantom{aaa} {\rm [strong-coupling]}.
\label{strong}
\ee
These results are indicated with ${\rm x}$'s, and only for $V=10t$ in Fig.~(\ref{fig1}b), where it is seen to be very accurate. Here we caution the reader
that it is an accurate approximation to the fully self-consistent solution as indicated, but in fact BCS theory itself is not expected to be very accurate
in this regime at finite temperature. So here it merely serves as a check that our solutions to the equations are accurate. 

The opposite case, that of weak coupling, is the one most normally used; furthermore we expect BCS theory to be reasonably accurate, at least in
three dimensions. In two dimensions, these results are also generally not so accurate, because Kosterlitz-Thouless physics\cite{kosterlitz73} is expected to
come into play. Our approximation follows the standard one,\cite{tinkham96} but accommodates the density of states with the logarithm singularity
[Eq.~(\ref{dense_2d_app})]; we obtain
\be
{T_c \over \omega_D} \approx 1.134 \ {\rm exp}\biggl\{A - \sqrt{A^2 + {4 \pi^2 t \over V} - B - {\rm log}[2(1.134)]} \biggr\},
\label{weak}
\ee
(weak coupling) where
\be
A \equiv A(\omega_D/t) \equiv 1 + {\rm log}\left( {8t \over 1.134 \omega_D}\right)
\label{a_defn}
\ee
and 
\be
B \equiv \int_0^\infty  dx \ {\rm sech}^2 x \ {\rm log}^2 x \ \ \approx \ 1.989....
\label{b_defn}
\ee
These results are indicated with asterisks for $V=2t$ and $V=1t$ in Fig.~(\ref{fig1}b), where they are indeed very accurate. Although
not shown, they become less accurate as $V$ increases. Notice, however, that even for $T_c/\omega_D \approx 0.4$ ($V=2t$ and $\omega_D \rightarrow 0$)
the weak coupling approximation works very well. Eq.~(\ref{weak}) is useful to illustrate how the logarithmic singularity in the density 
of states changes the usual exponential suppression for superconducting $T_c$ to one that can be significantly enhanced, as now the {\it square root}
of the inverse coupling strength appears in the exponential, a fact first pointed out, to our knowledge, in Ref.~[\onlinecite{hirsch86}].
This is most readily seen by allowing $V/t \rightarrow 0$ in Eq.~(\ref{weak}), to get
\be
{T_c \over \omega_D} \approx 1.134 \ {\rm exp}\biggl\{- \sqrt{ {4 \pi^2 t \over V}} \biggr\}.
\label{weakweak}
\ee

\subsection{Beyond Nearest Neighbour Hopping}

\begin{figure}[tp]
\begin{center}
\includegraphics[height=3.2in,width=3.0in,angle=-90]{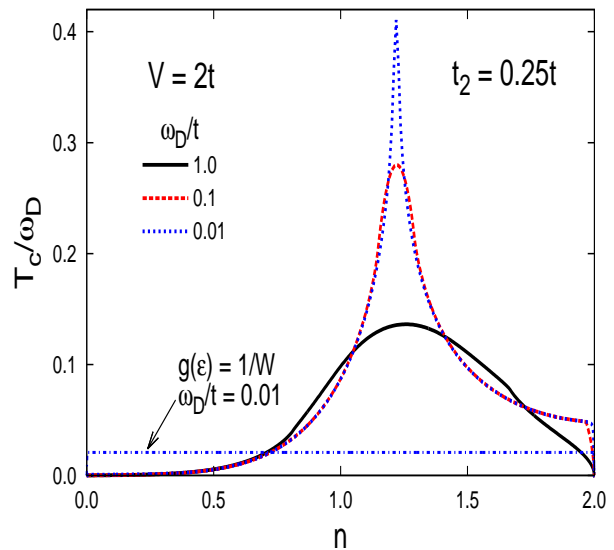}
\end{center}
\caption{Plot of $T_c/\omega_D$ vs. $n$ for $V/t = 2$, and 
$\omega_D/t = 1.0, 0.1, 0.01$, for the 2D case where next-nearest-neighbour (NNN) hopping is also present. The van Hove
singularity is now located at $\epsilon = 4t_2$ which corresponds to a filling $n \approx 1.2$. Also shown is the result for
a constant density of states, $g(\epsilon) = 1/W$, where $W = 8t$ is the electron bandwidth. The same enhancement occurs as
when NNN hopping is not included, when the chemical potential approaches the energy of the van Hove singularity. As was the
case in Fig.~1 the enhancement of $T_c/\omega_D$ is amplified as $\omega_D$ decreases.}
\label{fig2}
\end{figure}
Going beyond nearest neighbour hopping in two dimensions destroys the particle hole symmetry, but the singularity in the density of states
remains, albeit at some different value for the chemical potential (i.e. filling). The Appendix displays the Density of States for various
values of the next-nearest-neighbour (NNN) hopping parameter. Fig.~\ref{fig2} shows $T_c/\omega_D$ vs filling and again illustrates
that a significant enhancement occurs when the chemical potential is close to the van Hove singularity.

\subsection{Isotope Effect}

The partial isotope coefficient is defined by\cite{marsiglio08}
\be
\beta_i \equiv -{d {\rm ln} T_c \over d{\rm ln} M_i}
\label{iso_coeff}
\ee
where $M_i$ is the mass of the $i^{th}$ element; the total isotope coefficient ($\beta$) is the sum of these, and for the purpose of this
work we will assume an elemental superconductor; furthermore, for the harmonic approximation $\omega_D \propto 1/\sqrt{M}$,
Eq.~(\ref{bcs_weak_tc}) implies the expected standard result, $\beta = 1/2$. This positive value indicates that increasing the Debye
frequency is expected to raise $T_c$.

\begin{figure}[tp]
\begin{center}
\includegraphics[height=2.7in,width=2.1in,angle=-90]{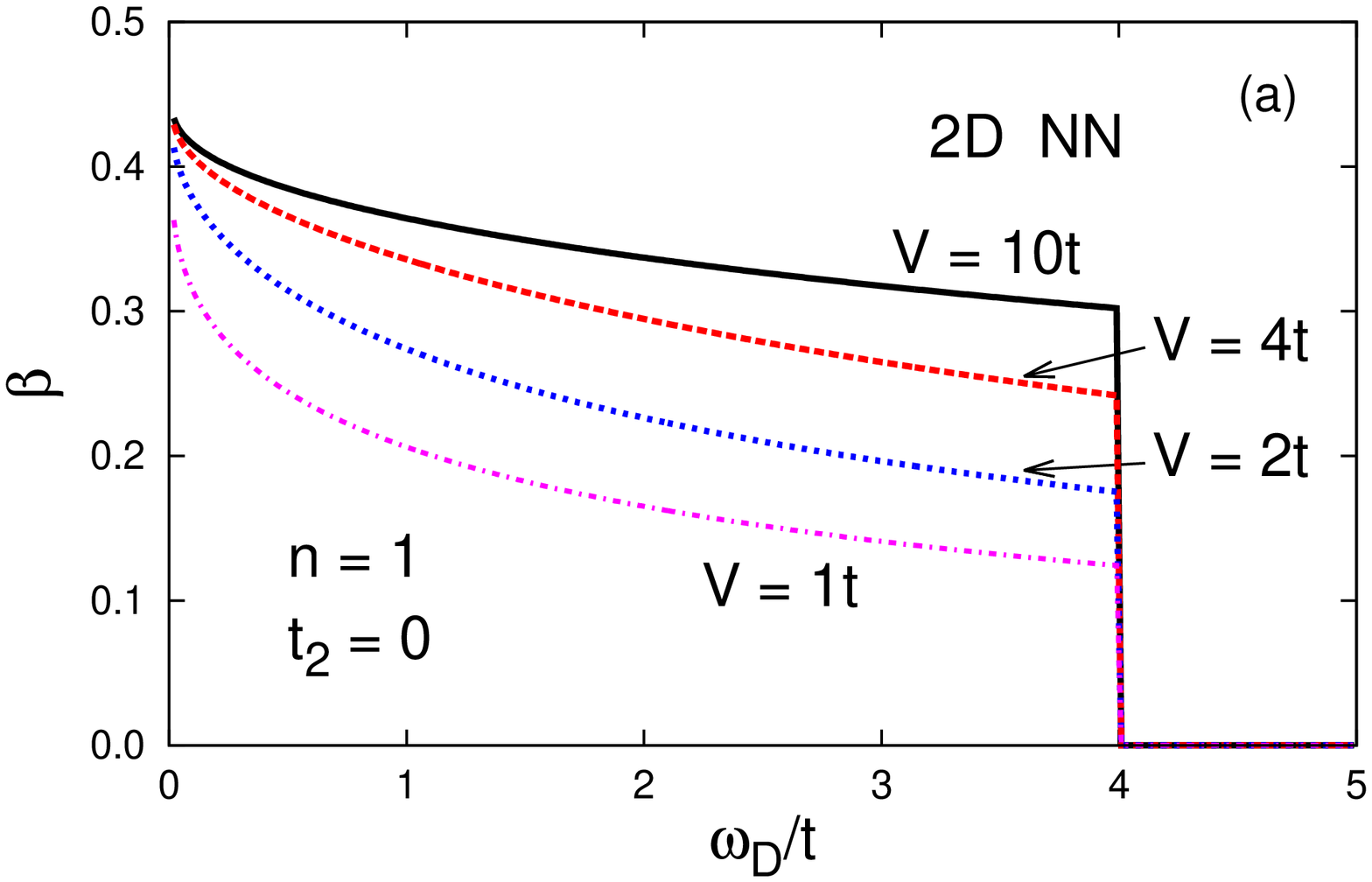}
\includegraphics[height=2.7in,width=2.1in,angle=-90]{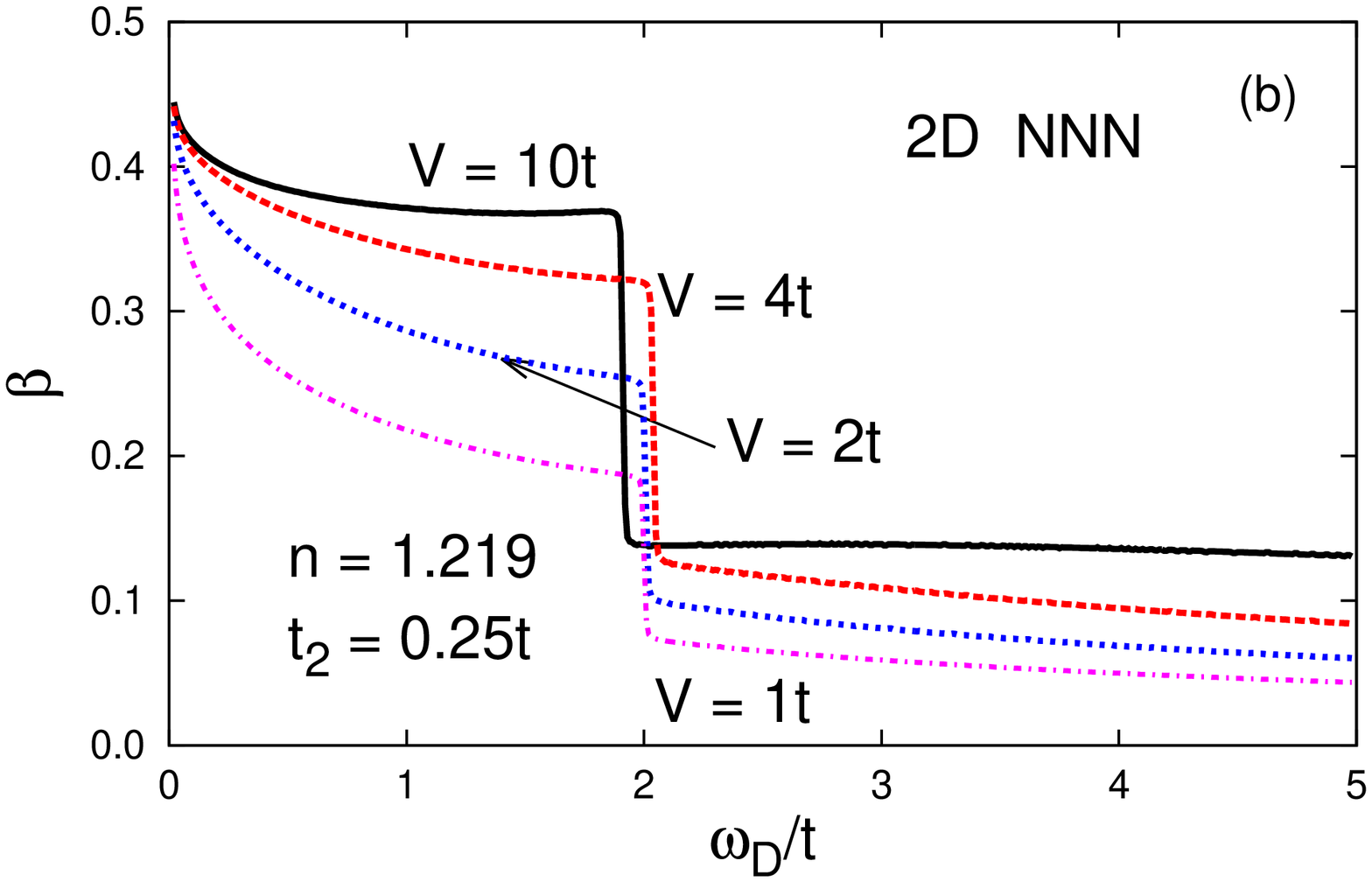}
\includegraphics[height=2.7in,width=2.1in,angle=-90]{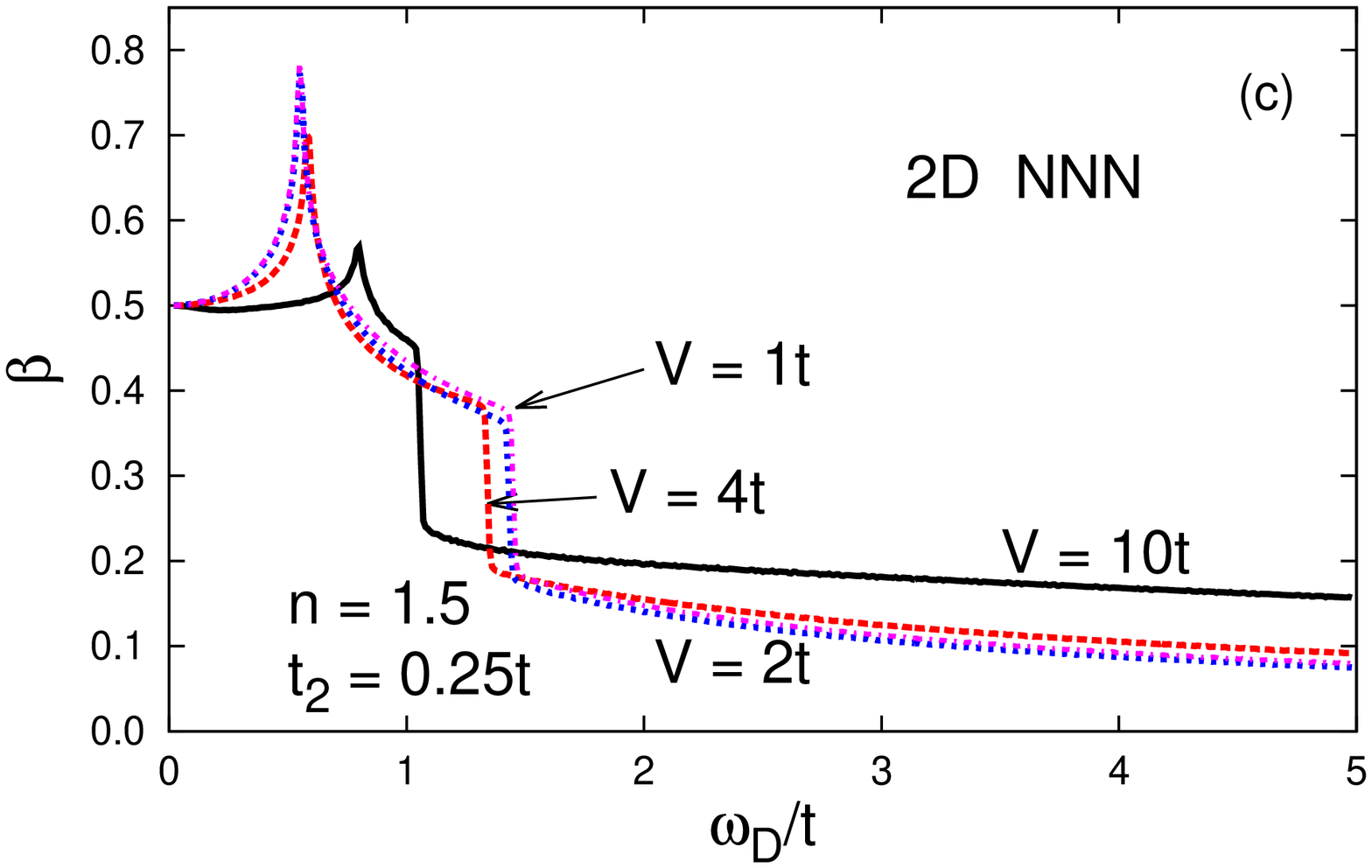}
\end{center}
\caption{Isotope coefficient $\beta$ vs $\omega_D/t$ for (a) NN hopping only and $n=1$, (b) NNN hopping with $t_2=0.25t$
and $n=1.219$, and (c) NNN hopping, again with $t_2 = 0.25t$, but with $n=1.5$. In (a) and (b) the filling is such that the 
chemical potential is at the van Hove singularity. In both these cases the behaviour is qualitatively similar and the
isotope coefficient decreases as a function of $\omega_D/t$ (Note that the physically relevant regime is $\omega_D/t << 1$).
Moreover, the most significant decrease occurs in both instances for weaker coupling, which is anyways where BCS theory
is to be most trusted. In (c) the chemical potential is away from the van Hove singularity. This results in a peak in the
isotope coefficient, at a value of $\omega_D$ which tracks the energy difference between the chemical potential and the
van Hove singularity. This is as expected, as the highest impact on $T_c$ will occur when the the value of $\omega_D$ can
include the states at the van Hove singularity. At large (and unrealistic) values of $\omega_D$ the isotope coefficient decreases
significantly as two (a) or one (b and c) band edges replaces $\omega_D$ as the cutoff.}
\label{fig3}
\end{figure}

The presence of a non-constant density of states will quantitatively change this result; in particular, if the chemical potential 
is at a van Hove singularity, then decreasing the mass (i.e. increasing the Debye frequency) will increase $T_c$ less than what
one would expect normally. This is because more states are included in the energy-lowering due to condensation, as before, but
the energy regime where this occurs (about $\omega_D$ on either side of the chemical potential) has a lower electronic density of
states, so the incremental benefit is decreased from what it is if the density of states is constant.

Fig.~\ref{fig3} shows the isotope coefficient vs. $\omega_D/t$ for (a) nearest-neighbour (NN) hopping only, at half-filling, 
(b) NNN with the chemical potential at the van Hove singularity ($n=1.219$), and (c) NNN with a filling of $n = 1.5$. 
The results in Fig.~\ref{fig3}(a) and (b) are qualitatively similar; the coefficient rapidly decreases as $\omega_D$ increases,
as would be expected, since the relevant energy regime moves further away from the singular part of the density of states with
increasing $\omega_D$. In both cases a precipitous drop occurs when $\omega_D$ exceeds a value that corresponds to the
distance from the chemical potential to the band edge. While these values are unrealistically large, it is worth understanding
what is occurring. In the case of NN hopping this value is $4t$, and then the isotope coefficient becomes zero, since further
increasing $\omega_D$ plays no role in determining $T_c$, as the role of the cutoff is now taken by the band edge, and not
$\omega_D$. For NNN hopping only one band edge takes on the role of the cutoff; the second bandwidth will enter for
larger values of $\omega_D/t$ than those shown.

In  Fig.~\ref{fig3}(c) the chemical potential is well away from the van Hove singularity; then the isotope coefficient $\beta$
peaks in value when the value of $\omega_D$ is `tuned' to equal the difference in energy between the chemical potential
and the energy of the van Hove singularity. Coupling to these states has the most significant effect on $T_c$, which results
in a peak and in the achievement of anomalously high values of $\beta$.

\subsection{The zero temperature energy gap}

\begin{figure}[tp]
\begin{center}
\includegraphics[height=3.4in,width=3.0in,angle=-90]{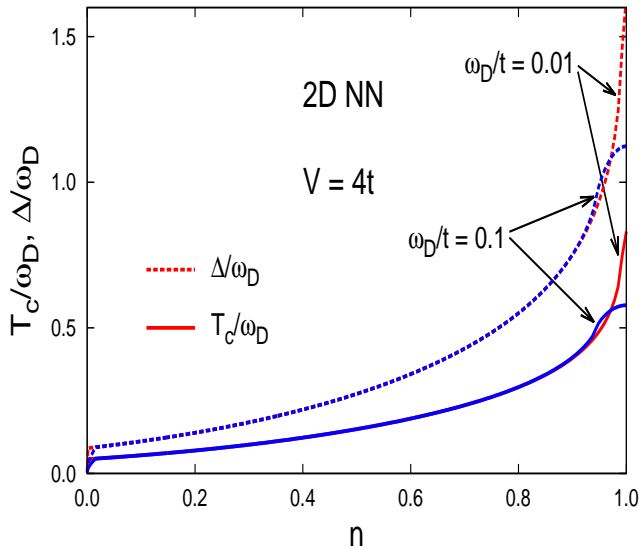}
\end{center}
\caption{$T_c$ and the zero temperature order parameter, $\Delta$, vs. electron density $n$, for the 2D case with NN hopping
only, with $V=2t$ and $\omega_D = 0.1t$. The behaviour of $\Delta$ follows
that of $T_c$, with a slight enhancement of the ratio as the singularity is approached.}
\label{fig4}
\end{figure}

One can ask if the just-described effects of a van Hove singularity similarly apply to the finite temperature energy gap.
We have thus solved the finite temperature gap equations, Eqs.~(\ref{bcs1a},\ref{bcs2a}) in several representative cases. For the most part
we find that little differs for the pairing gap, $\Delta$, as a function of vicinity of the Fermi energy to the van Hove singularity.
For example, the temperature dependence, $\Delta (T)$, as a function of temperature $T$ is barely discernible from the usual
temperature dependence obtained with a constant density of states. 
By way of example, we show in Fig.~\ref{fig4} the pairing gap at zero temperature, $\Delta$, and the superconducting critical temperature,
$T_c$, as a function of electron density, $n$, for the 2D case with NN hopping only. Both peak near $n=1$, i.e. the location of the
van Hove singularity in the density of states; in this sense, $\Delta$ tracks $T_c$. The ratio for example, $2\Delta/(k_BT_c)$, grows
from 3.53 at low densities to about 3.7 at half-filling, a rather insignificant change. We turn to three dimensions now, and focus on $T_c$.

\section{Three Dimensions}

We now turn to similar calculations in 3D. As summarized in Eqs.~(\ref{sc}-\ref{fcc}), we focus only on the cubic lattice structures, sc,
fcc, and bcc. The densities of states for these were first calculated by Jelitto\cite{jelitto69} and are provided in the Appendix for cases
involving NNN hopping as well. \red{As we mentioned earlier, while these calculations were performed almost half a century ago,
most researchers are not aware\cite{remark2} that singularities indeed exist in the tight-binding model for the bcc and fcc cases (for NN hopping only) and
even in the sc case when NNN hopping is included. This is true only for `special' values of the hopping parameters, but as detailed in
the appendix,} remnants of these singularities remain even when \red{other values of the hopping parameters are used.} 
Based on what we found in two dimensions, along with some exploratory calculations, here we will focus
on $T_c$ and the isotope coefficient, $\beta$; results for $\Delta$ follow those of $T_c$, as we found in two dimensions.

\subsection{simple cubic NN}

\begin{figure}[tp]
\begin{center}
\includegraphics[height=3.8in,width=3.2in,angle=-90]{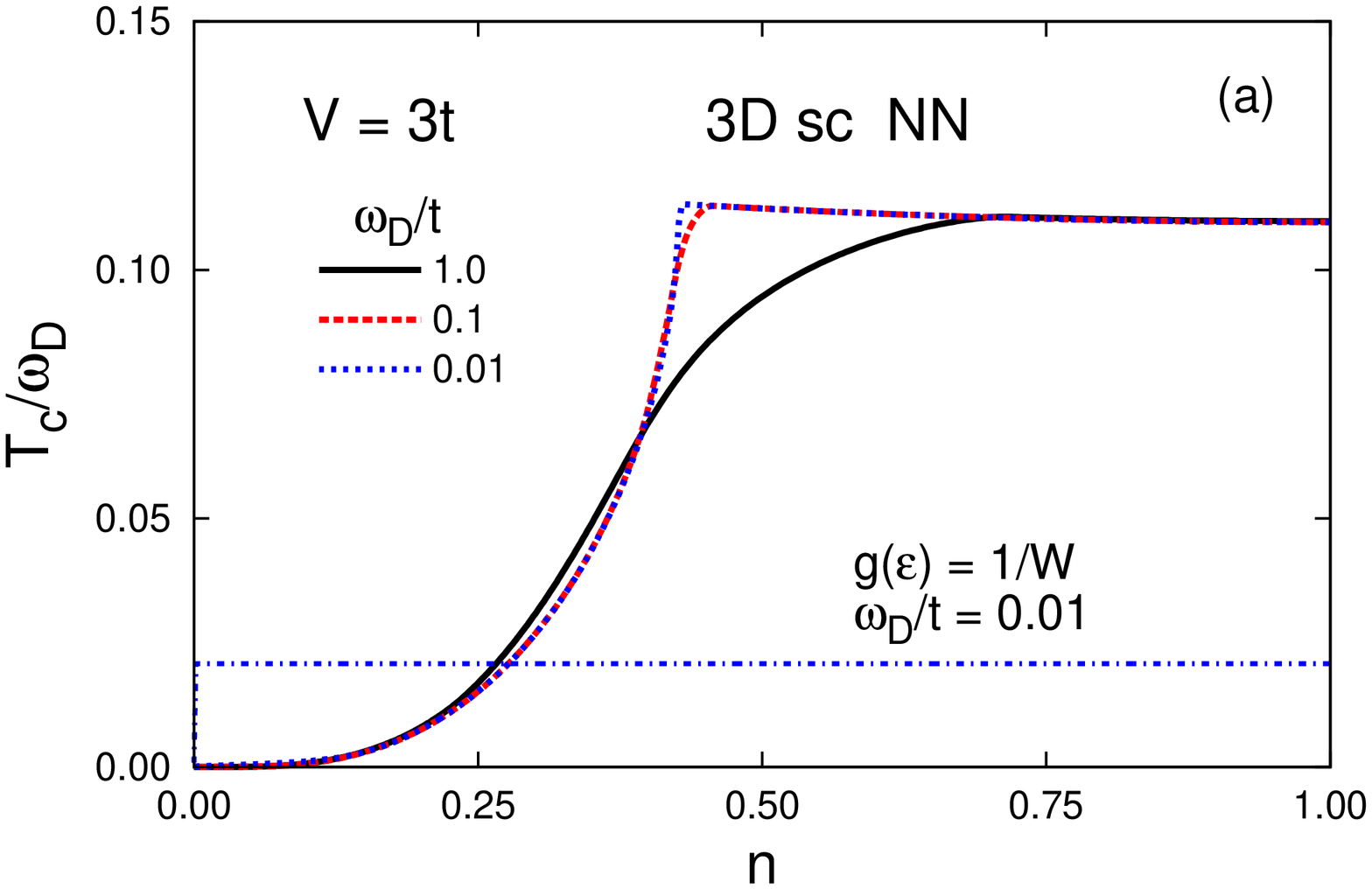}
\includegraphics[height=3.1in,width=2.5in,angle=-90]{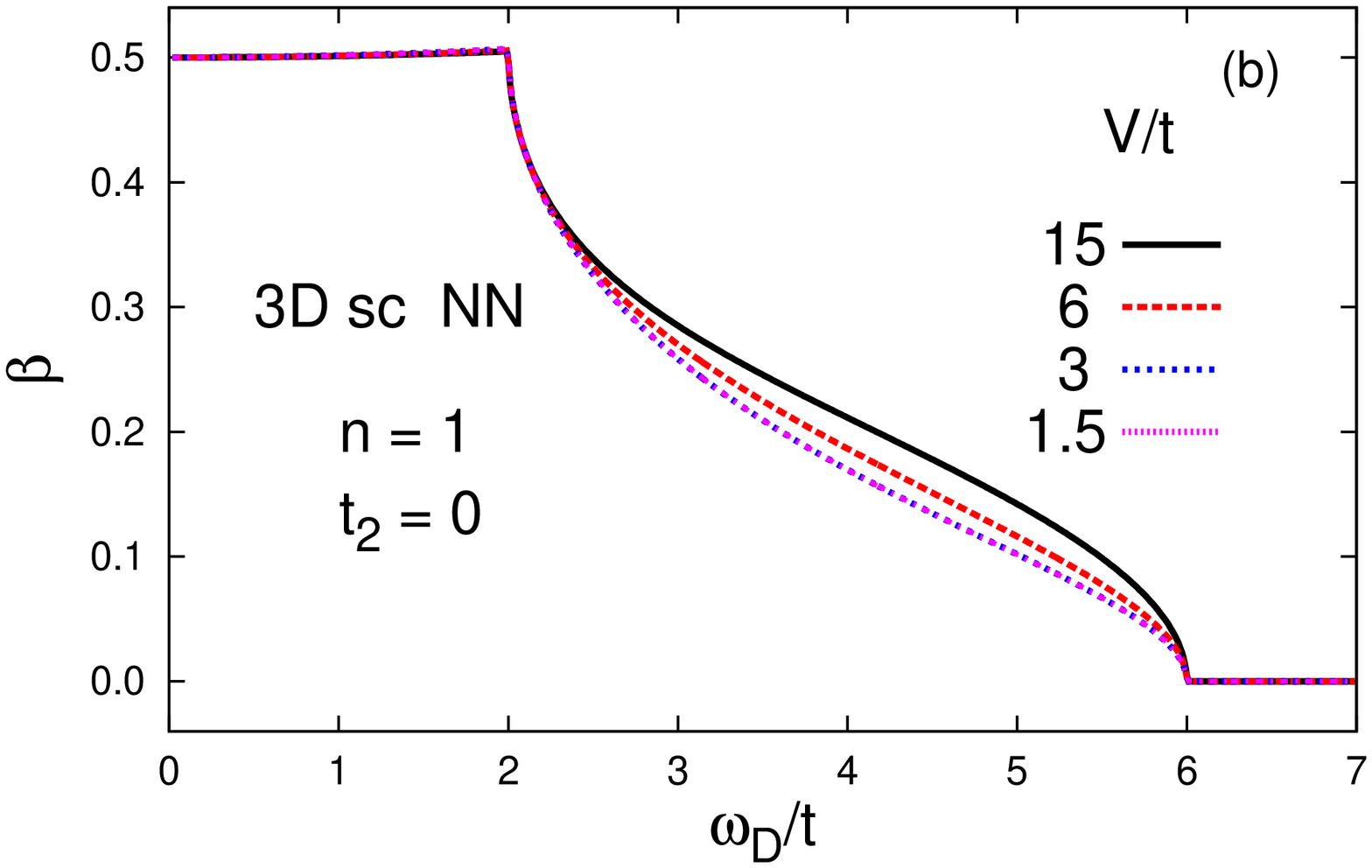}
\end{center}
\caption{(a) $T_c$ vs. filling, $n$ and (b) the isotope coefficient $\beta$ vs $\omega_D/t$ for a variety of values of $\omega_D$
in (a) and for a number of coupling strengths in (b). These results are for the simple cubic three-dimensional case, with NN hopping only, and
for $n=1$. Results are as expected and as explained in the text.}
\label{fig5}
\end{figure}

The simple cubic density of states consists of van Hove singularities only in the derivative of the density of states with respect
to energy (see the red curve in Fig.~\ref{figAsc} in the Appendix). The behaviour  of $T_c$ is therefore not so unusual. Fig.~\ref{fig5}(a) shows $T_c/\omega_D$ vs. $n$
for a fairly weak coupling case ($V = 3t$) from zero density to half-filling ($n=1$). Note that this lattice is bipartite and has particle-hole symmetry.
Hence, results for $n > 1$ are a mirror reflection of those for $n < 1$, and we display only the latter.
We show results for three values of $\omega_D$; in fact as long as
$\omega_D << t$ the electron density of states at the chemical potential plays the most important role, as is evident from how
$T_c$ tracks $g(\epsilon_F)$, albeit as a function of occupation rather than as a function of energy. Only for $\omega_D = t$ does the
$T_c$ curve begin to become ``rounded'' compared to the density of states. Also shown is the result obtained for a constant density of
states, $1/W$, where $W = 12t$ for the three dimensional simple cubic tight-binding model.  In Fig.~\ref{fig5}(b) we show the isotope coefficient
as a function of $\omega_D/t$ for four different values of the coupling strength, $V/t$. The results for $V=3t$ and $V=1.5t$ cannot be
distinguished from one another, indicating that $V=3t$ is already in the weak coupling limit. The isotope coefficient becomes reduced from the
`canonical' value of $0.5$ only when $\omega_D$ increases beyond the energy of the first van Hove singularity near the origin, at $\pm 2t$.
The $\beta$ decreases steadily to zero, achieved for $\omega_D \ge W/2 = 6t$. The dependency on coupling strength is very minor.
Away from half-filling there are no surprises, and both $T_c$ and $\beta$ track the density of states at the chemical potential. As was the case
in two dimensions, the isotope coefficient displays a peak when the size of $\omega_D$ allows a coupling to states with significantly higher
density of states, i.e. when $|\mu| > 2t$, where $\mu$ is the chemical potential.
 
\subsection{simple cubic NNN}

\begin{figure}[tp]
\begin{center}
\includegraphics[height=3.7in,width=3.3in,angle=-90]{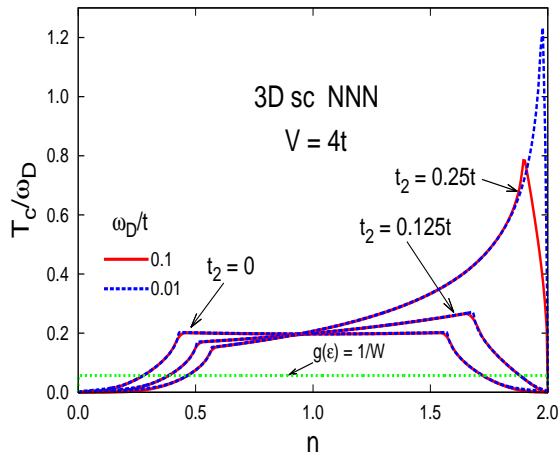}
\end{center}
\caption{Plot of $T_c/\omega_D$ vs. $n$ for $V/t = 4$, and 
$\omega_D/t = 0.1, 0.01$, for three different values of $t_2/t = 0, 0.125, 0.250$. Note that the results are relatively insensitive to $\omega_D$ 
except for $t_2 = 0.25t$, where a singularity exists in the density of states near the top of the band (see Fig.~\ref{figAsc} in the Appendix), and
$T_c/\omega_D$ continues to increase near $n=2$ as $\omega_D$ decreases. Also shown is the result for
a constant density of states, $g(\epsilon) = 1/W$, where $W = 12t$ is the electron bandwidth for the sc lattice with $|t_2|/t \le 1/4$. 
This latter result is not sensitive to $\omega_D$ except at the band edges, and is shown only for $\omega_D/t = 0.01$.}
\label{fig6new}
\end{figure}

Remarkably, including sufficient NNN hopping in the sc lattice results in a singularity in the density of states at the top of the band
(see the blue curve in Fig.~\ref{figAsc} in the Appendix), similar to what occurs for the FCC lattice with NN hopping only (see below). In Fig.~\ref{fig6new}
we show $T_c/\omega_D$ vs. electron density for three different values of the NNN hopping, $t_2/t = 0, 0.125, 0.250$ and two different
values of $\omega_D$. The value of $\omega_D$ is not so important except in the case $t_2/t = 0.25$, where a singularity occurs in the
density of states; this results in a singularity in $T_c/\omega_D$ near $n=2$ as $\omega_D$ decreases. Also shown is the result for a constant
density of states with value $g(\epsilon) = 1/W$, where $W = 12t$ is the electron bandwidth for the sc lattice with $|t_2|/t \le 1/4$. Clearly the 
potential enhancement of $T_c/\omega_D$ is very large at high fillings. A particle-hole symmetry exists with these results for negative values
of $t_2/t$ (not shown). The important point is that for values of $t_2/t$ close to $0.25$ a peak will remain in the density of states, giving
rise to a large enhancement in $T_c/\omega_D$.

\subsection{body-centred cubic NN}

\begin{figure}[tp]
\begin{center}
\includegraphics[height=3.7in,width=3.3in,angle=-90]{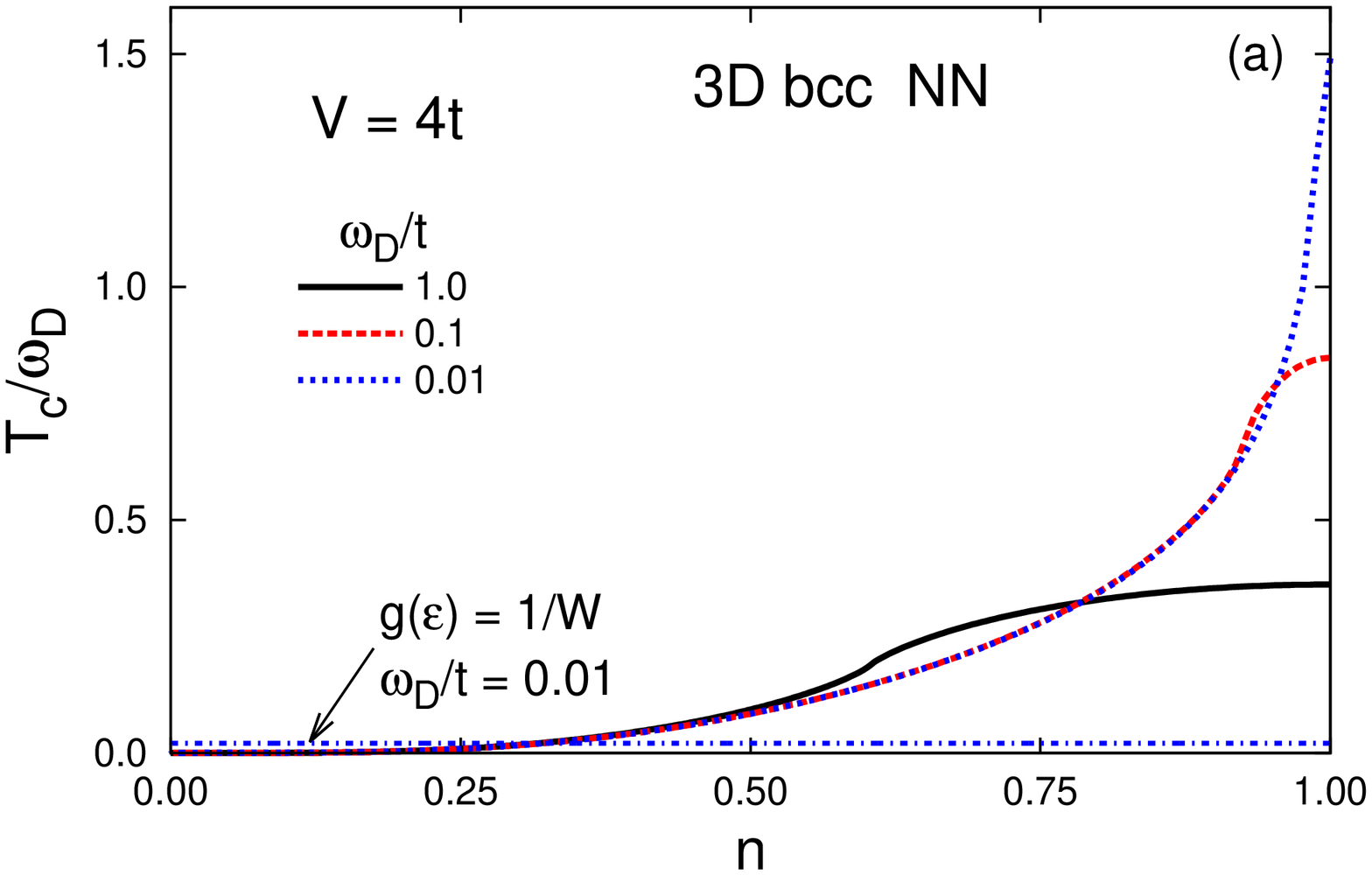}
\includegraphics[height=3.0in,width=2.6in,angle=-90]{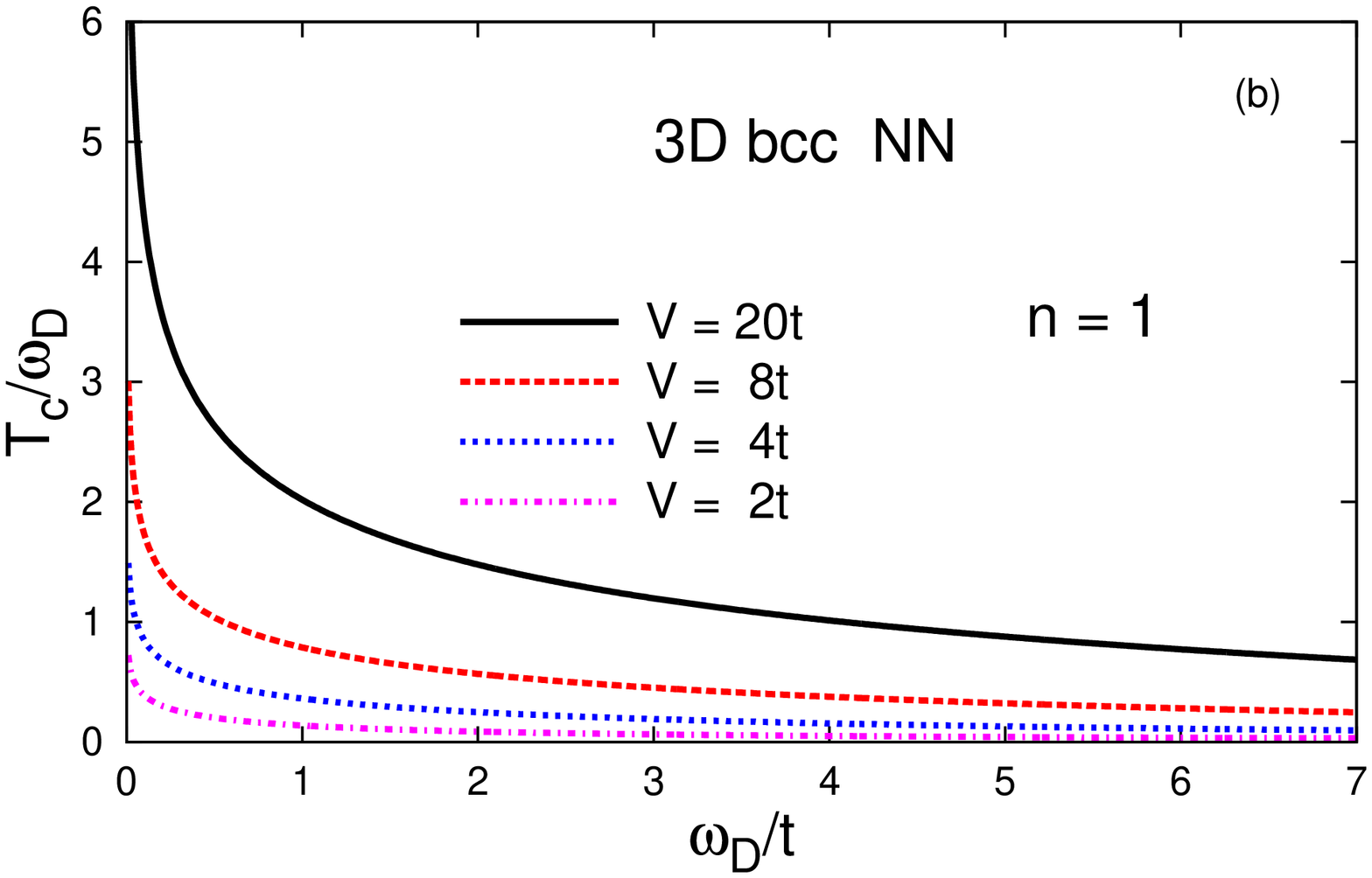}
\end{center}
\caption{(a) Plot of $T_c/\omega_D$ vs. $n$ for $0 < n < 1$ (the results for $2 > n > 1$ are symmetric) for $V/t = 4$, and 
$\omega_D/t = 1.0, 0.1, 0.01$. Also shown is the result for
a constant density of states, $g(\epsilon) = 1/W$, where $W = 16t$ is the electron bandwidth. This latter result is not sensitive to $\omega_D$ 
except at the band edges. There is a significant enhancement near the van Hove singularity, which continues to grow without bound for decreasing
$\omega_D/t$. (b) $T_c/\omega_D$ vs. $\omega_D/t$ for $n=1$ for various values of $V$. This view highlights the sharp increase
in $T_c/\omega_D$ as $\omega_D \rightarrow 0$. }
\label{fig6}
\end{figure}

As noted in the Appendix the tight-binding model with a bcc lattice with NN hopping only displays a singularity at $\epsilon = 0$. This singularity
is a logarithm squared and hence stronger than the two-dimensional singularity which diverges logarithmically. Fig.~\ref{fig6}(a) 
shows $T_c/\omega_D$ vs. $n$ for a fairly weak coupling case ($V = 4t$) for zero density to half-filling ($n=1$). Like the sc 3D case, the bcc lattice
is bipartite, and with NN hopping only, this lattice has particle-hole symmetry. Hence, as in that case, 
results for $n > 1$ are a mirror reflection of those for $n < 1$, and we display only the latter.
We show results for three values of $\omega_D$; again, as long as
$\omega_D << t$, the electron density of states at the chemical potential plays the most important role. In particular, for the smallest value
of $\omega_D$ shown, $T_c$ again tracks $g(\epsilon_F)$ as a function of occupation (rather than as a function of energy).
The enhancement above the result for a constant density of states with the same bandwidth (horizontal line just above zero) is
enormous. Here, $W = 16t$ for the bcc NN tight-binding model.
In Fig.~\ref{fig6}(b) we show the same quantity,  $T_c/\omega_D$ vs. $\omega_D$ for a variety of coupling strengths at half-filling. As expected, this
BCS calculation shows $T_c$ increasing with $V$; we remind the reader again that this calculation is expected to be valid only for some weak coupling
range. The important point is that $T_c/\omega_D$ eventually diverges as $\omega_D$ decreases, because the density of states at $\mu = 0$ is
diverging, and the density of states right at the Fermi level becomes the only key quantity as the Debye frequency decreases.

\begin{figure}[tp]
\begin{center}
\includegraphics[height=3.4in,width=3.0in,angle=-90]{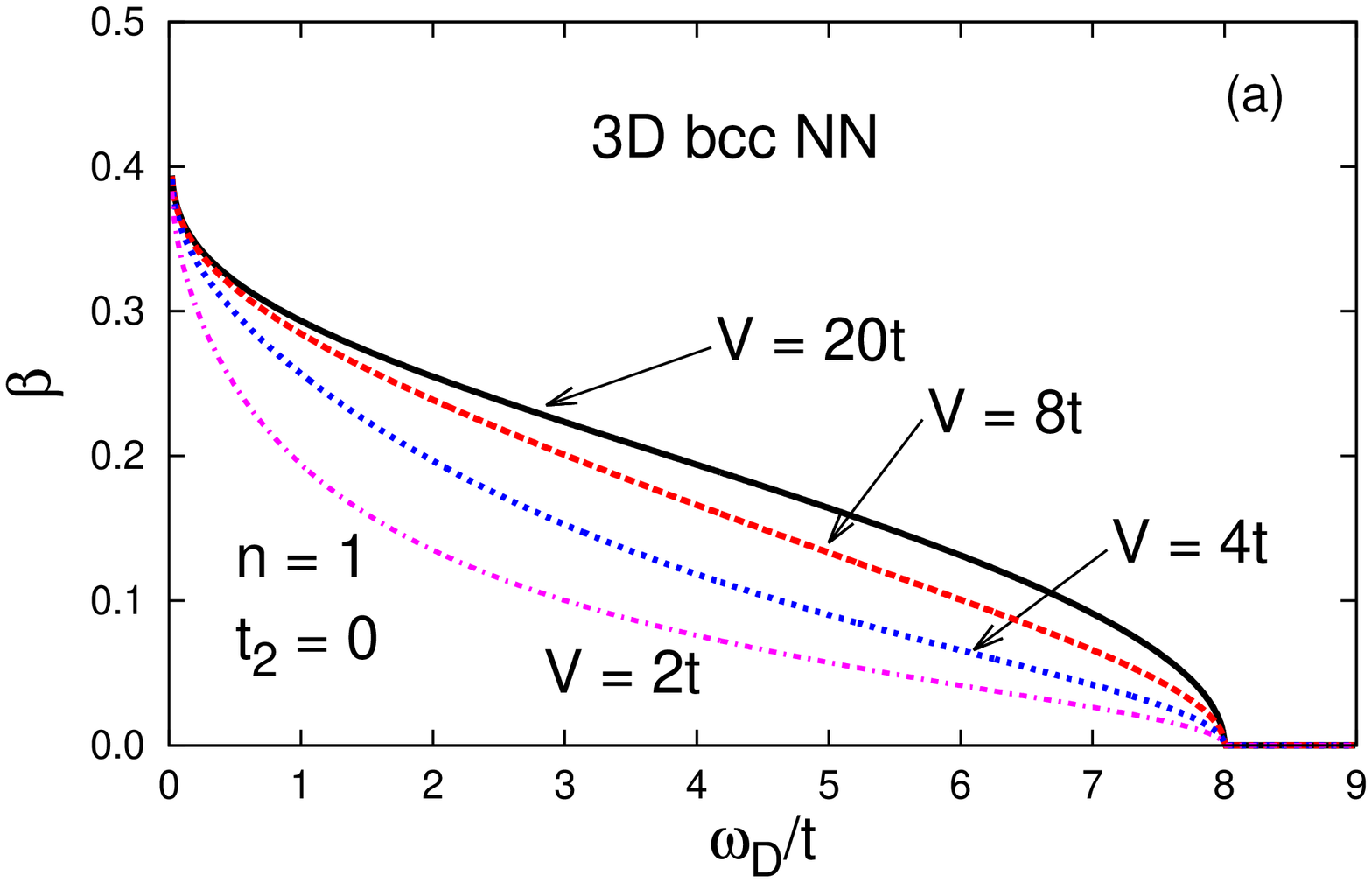}
\includegraphics[height=3.4in,width=3.0in,angle=-90]{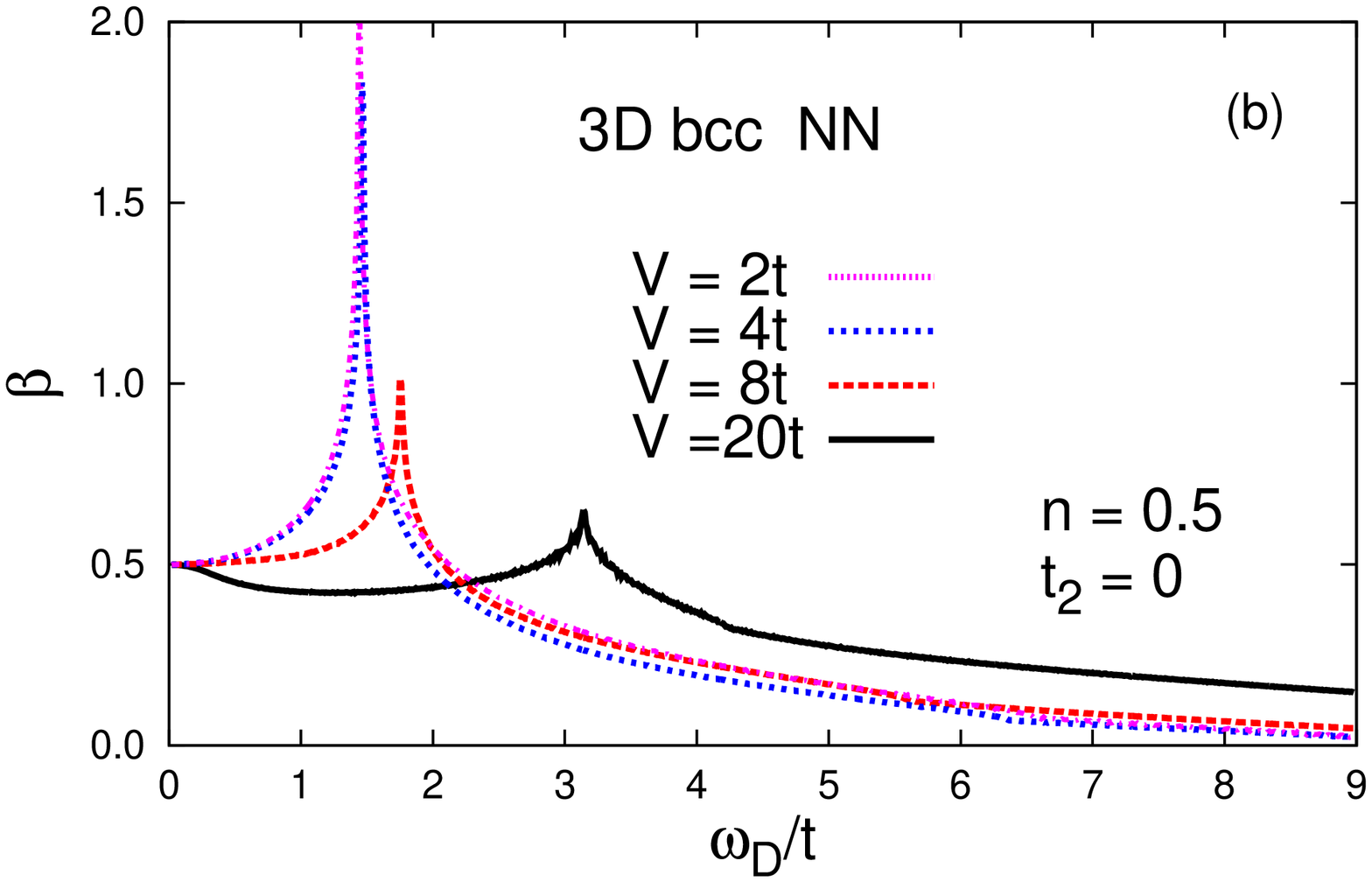}
\end{center}
\caption{(a) Isotope coefficient $\beta$ vs. $\omega_D/t$ for the 3D bcc NN case ($t_2 = 0$), at half-filling, for a variety of coupling strengths. The
isotope coefficient is significantly reduced from $0.5$, due to the singularity in the density of states. (b)  Isotope coefficient $\beta$ vs. $\omega_D/t$ for the 3D bcc NN case ($t_2 = 0$), at quarter filling $n=0.5$, for the same coupling strengths as in (a). The isotope coefficient now has a significant
peak, particularly for weak coupling, when $\omega_D$ is such that states in the peak of the density of states are primarily included in determining
$T_c$ (see text).}
\label{fig7}
\end{figure}

Away from half-filling results are again as expected; the chemical potential is at an energy where the density of states is relatively low. As $\omega_D$
increases to a point where the singularity in the density of states becomes relevant then $T_c/\omega_D$ will peak. Unlike Fig.~\ref{fig6}(b), where
$T_c/\omega_D$ monotonically decreases as $\omega_D$ increases, $T_c/\omega_D$ is non-monotonic, i.e. the result is sensitive to the singularity
{\it not at} the Fermi level.

The isotope coefficient is similar to what we have seen before; in Fig.~\ref{fig7}(a) we show $\beta$ as a function of $\omega_D/t$ for $n=1$ and
for a variety of coupling strengths. Clearly the isotope coefficient is greatly reduced even for fairly low values of $\omega_D/t$. In Fig.~\ref{fig7}(b)
we show the same result for $n=0.5$; now the isotope coefficient peaks to very high values, as values of $\omega_D$ are reached that bring the
singularity in the electron density of states in ``resonance'' with the chemical potential through $\omega_D$. That is, a prominent peak in $\beta$
occurs, particularly in weak coupling, when $\mu + \omega_D = \epsilon_{\rm sing}$, where $\epsilon_{\rm sing} = 0$ is the energy at which the
singularity occurs. We thus have the intriguing possibility of an anomalously high isotope coefficient associated with a non-optimal critical temperature.

The bcc tight-binding model shows significant enhancement for $T_c$, for a wide range of electron density (Fig.~\ref{fig6}(a)). However, as was the
case in 2D, it is important to examine the impact of a NNN hopping parameter. We do this in the next subsection.

\subsection{body-centred cubic NNN}

\begin{figure}[tp]
\begin{center}
\includegraphics[height=3.8in,width=3.4in,angle=-90]{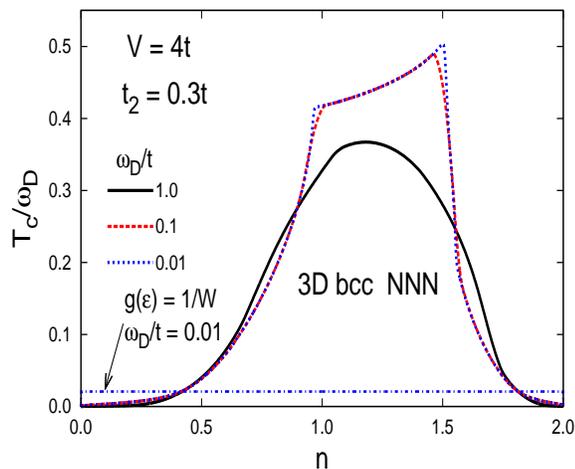}
\end{center}
\caption{Plot of $T_c/\omega_D$ vs. electron density $n$ for $V/t = 4$, for the BCC lattice structure, now with NNN hopping, $t_2 = 0.3t$, and 
$\omega_D/t = 1.0, 0.1, 0.01$. Also shown is the result for
a constant density of states, $g(\epsilon) = 1/W$, where $W = 16t$ is the electron bandwidth. As before, this latter result is not sensitive 
to $\omega_D$ except at the band edges. There is a clear enhancement of $T_c$ over a wide range of densities near the maximum
in the Density of States (see Fig.~\ref{figA2}). Note, however, that because the density of states is no longer singular, $T_c/\omega_D$
now saturates as $\omega_D$ decreases (red to blue curve).}
\label{fig8}
\end{figure}

The introduction of NNN hopping for the BCC lattice structure changes the nature of the density of states in a profound way. As illustrated in the Appendix,
the singular behaviour is entirely removed. Nonetheless, a highly peaked structure remains, which we expect will continue to cause a considerable
enhancement of $T_c$. This enhancement is more significant than the one found at half-filling since the Fermi surface is no longer nested, and therefore
competing instabilities to superconductivity will be suppressed. Since the density of states is not singular, however, we expect that $T_c$ will {\it not}
continue to increase as $\omega_D$ is decreased (as was the case in Fig.~\ref{fig6}(a) at $n=1$). Nonetheless, as noted in the Appendix, the maximum
is, in many respects, more `robust' than in the NN case, in that a larger area is contained in the maximum region than in the NN case. This was
also the case in 2D, and is the case for the FCC lattice structure (see Fig.~\ref{figA3} below).

Figure~\ref{fig8} shows $T_c/\omega_D$ vs. $n$, for a variety of values of $\omega_D$, for the particular case of $t_2= 0.3t$. As anticipated,
the enhancement with decreasing $\omega_D$ now saturates; near the peak values, $T_c/\omega_D$ hardly increases as $\omega_D = 0.1t$
(red curve) decreases to $\omega_D = 0.01t$ (blue curve). This is in contrast to the scenario shown in Fig.~\ref{fig6}, where, near the peak 
electron density, $T_c/\omega_D$ continues to increase indefinitely as $\omega_D$ decreases. Here, however, the Fermi surface is no 
longer nested, and competing instabilities, not considered here, will be significantly suppressed (see, for example, Fig.~1 in 
Ref. [\onlinecite{marsiglio90a}] for a demonstration of the suppression of a charge-density-wave instability due to the removal of nesting). Thus,
the large enhancement displayed in Fig.~\ref{fig8} will likely remain even when competing instabilities are considered.

\subsection{face-centred cubic NN and NNN}

\begin{figure}[tp]
\begin{center}
\includegraphics[height=3.8in,width=3.4in,angle=-90]{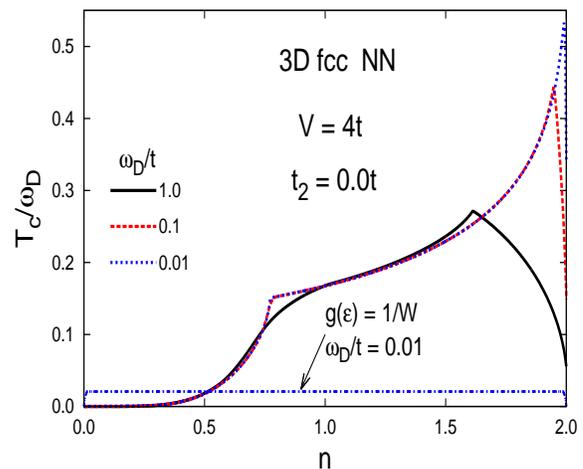}
\end{center}
\caption{Plot of $T_c/\omega_D$ vs. electron density $n$ for $V/t = 4$, for the FCC lattice structure, with NN hopping only, and 
$\omega_D/t = 1.0, 0.1, 0.01$. Also shown is the result for
a constant density of states, $g(\epsilon) = 1/W$, where $W = 16t$ is the electron bandwidth. As before, this result is not sensitive 
to $\omega_D$ except at the band edges. There is a clear enhancement of $T_c$ over a wide range of densities near the maximum
in the Density of States (see Fig.~\ref{figA3}). Note that near $n=2$, this maximum will increase without bound as $\omega_D$ decreases.}
\label{fig9}
\end{figure}

\begin{figure}[tp]
\begin{center}
\includegraphics[height=3.8in,width=3.4in,angle=-90]{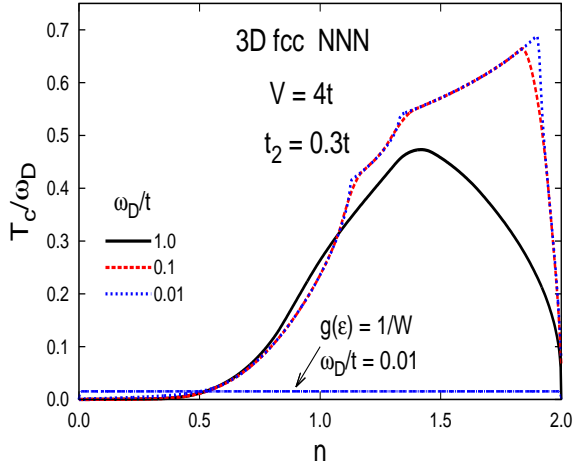}
\end{center}
\caption{
Plot of $T_c/\omega_D$ vs. electron density $n$ for $V/t = 4$, for the FCC lattice, now with NNN hopping, $t_2 = 0.3t$, and 
$\omega_D/t = 1.0, 0.1, 0.01$. Also shown is the result for
a constant density of states, $g(\epsilon) = 1/W$, where $W = 17.2t$ is the electron bandwidth for $t_2 = 0.3t$. Even for the non-constant
density of states results, note the lack of sensitivity of $T_c/\omega_D$ to $\omega_D$ over essentially all electron densities, 
for sufficiently small values of $\omega_D/t$.  There remains a clear enhancement of $T_c/\omega_D$ over a wide range of densities 
near the maximum in the Density of States (see Fig.~\ref{figA3}).}
\label{fig10}
\end{figure}

Similar to the BCC NN case, the electron density of states for an FCC lattice with NN hopping only, also displays a singularity, right at
the band edge, as shown by the red curve, i.e. the rightmost curve, in Fig.~\ref{figA3} in the Appendix. In this case the FCC lattice is
{\it not} bipartite, so no nesting occurs. In Fig.~\ref{fig9} we show results for relatively weak coupling, $V=4t$, as a function of electron
density, for several values of $\omega_D$. The expected enhancement in $T_c/\omega_D$ occurs near the singularity, now at the top
of the band, and this increases indefinitely as $\omega_D$ decreases. Note that for decreasing $\omega_D$, enhancement of
$T_c/\omega_D$ continues only for a limited electron density region near the singularity.

This is expected to be a robust result, in that there will not be a large enhancement of competing instabilities due to the lack of nesting.
On the other hand, one always expects a small amount of NNN hopping, and as this removes the singularity in the electron density
of states, one might well ask whether the ensuing enhancement of $T_c/\omega_D$ will also disappear. Fig.~\ref{figA3}, which also
displays the electronic density of states for non-zero values of $t_2/t$, illustrates that a ``robust'' peak remains. We focus now on
results for $t_2/t = 0.3$. Fig.~\ref{fig10} shows $T_c/\omega_D$ vs electron density for relatively weak coupling, and shows that 
 a strong enhancement of $T_c/\omega_D$  continues to occur near the peak structure in the density of states. In fact the values of $T_c/\omega_D$
 are comparable to (or even greater than!) those achieved (for electron densities $1.0 < n < 1.9$) even with $\omega_D = 0.01t$ in the case with $t_2 = 0$ (see Fig.~\ref{fig9}), where a singularity exists in the density of states.


\subsection{\red{H$_3$S as a case of bcc with NNN hopping}}

\begin{figure}[tp]
\begin{center}
\includegraphics[height=3.8in,width=3.4in,angle=-90]{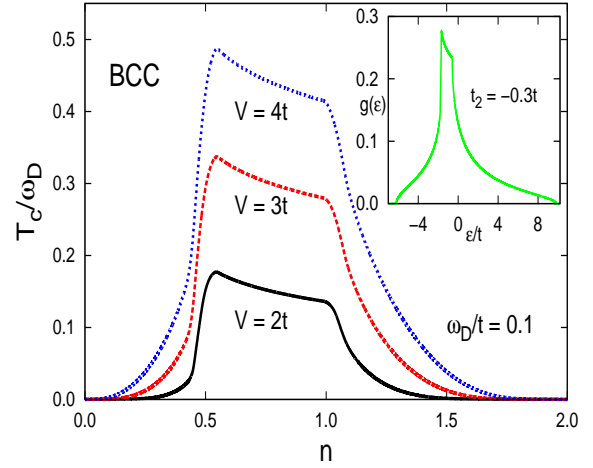}
\end{center}
\caption{\red{Plot of $T_c/\omega_D$ vs. electron density $n$ for  various values of coupling strength, $V/t = 2$, $3$, and $4$. We used the BCC lattice 
with NNN hopping, $t_2 = -0.3t$, which provides a reasonable facsimile to the peak in the density of states (see inset) calculated with DFT methods.
For illustration purposes, if  $V = 2t$ ($V/(16t) = 0.125$), with nearest-neighbour hopping $t = 1$ eV, and $\omega_D = 100$ meV, 
then $T_c \approx 160$ K (at $n=1$). In this range of $\omega_D$ the results for $T_c/\omega_D$ are insensitive to $\omega_D$, and therefore
$T_c$ scales with $\omega_D$.}}
\label{fig11}
\end{figure}

\red{First-principles calculations show that the Fermi energy occurs near a 
well-defined peak in the electronic density of states (see Fig.~8 of Ref. [\onlinecite{quan16}]
and Fig.~2b or 9 of  Ref. [\onlinecite{sano16}].
A very reasonable facsimile of this peak is given by our tight-binding model for the BCC lattice
structure with a negative NNN hopping parameter, as shown in the inset of Fig.~\ref{fig11}. Not surprisingly,
T$_c$ as a function of electron density will display the same peak structure, as illustrated in the main
part of Fig.~\ref{fig11} (compare with Fig.~\ref{fig8}). While it is not so useful to attempt an actual fit to
T$_c$ with such limited data and such a limited theoretical framework (see Ref. [\onlinecite{sano16}] for positive
steps in removing several of the limitations), for the sake of completeness, we show T$_c$ vs electron 
density for a number of weak coupling strengths. By way of illustration, for $V = 2t$ ($V/(16t) = 0.125$),
with nearest-neighbour hopping $t = 1$ eV, and $\omega_D = 100$ meV (all conservative values), then $T_c \approx 160$ K.
We expect actual reductions due to retardation and other effects,\cite{sano16} but these estimates suggest
that the possibility of an enhanced T$_c$ due to a peaked electronic density of states is quite realistic.}

\section{Summary}

In the context of BCS theory, we have examined the role of van Hove singularities in the electronic density of states on
the superconducting critical temperature and the isotope coefficient (and briefly the pairing gap) for various band structures
given by tight-binding models in two and three dimensions. We have adopted the simplest kind of pairing potential, an
attractive interaction with energy scale $\omega_D$, which gives rise to an order parameter with s-wave symmetry. While
the model follows the original BCS\cite{bardeen57} paper and is therefore suggestive of a phonon-mediated interaction, it
in fact has greater generality.

Many such models have been proposed for high temperature superconductors, and many calculations have been performed,
as documented in the references. However, here we have gone beyond the existing literature in two respects. First, we
have systematically treated the tight-binding models over all electron densities, with careful account of the BCS number
equation (our Eq.~(\ref{bcs2}) below $T_c$ or Eq.~(\ref{bcs2a}) at $T_c$), and we have utilized the tight-binding density of
states as given, whether for NN hopping only or with NNN hopping included as well. Secondly, we have performed calculations
for systems with van Hove singularities in three dimensions. While the simple cubic lattice structure is well known to produce a 
density of states without singularities (the van Hove singularities manifest themselves in cusps and the derivatives of the density 
of states), it is less appreciated that both the face-centred cubic and the body-centred cubic exhibit singularities in their
densities of states.\cite{jelitto69} It is also surprising that the sc lattice structure results in a singular density of states when the NNN
hopping $t_2$ in increased to $t/4$. Well before this value is reached the density of states exhibits a strong peak.

We have illustrated that these singularities can give rise to very large enhancements in $T_c$. Even in the case of the sc lattice,
when NNN hopping is included, significant enhancements of $T_c$ can occur. The bcc lattice is bipartite and
therefore nested. As is well known from other nested Fermi surface problems (though certainly less studied for the bcc lattice
in particular) other competing instabilities are expected to play an important role, and the BCS calculations provided here become
doubtful, as superconducting $T_c$ will often be suppressed. \red{For example, a CDW instability would certainly compete
in the case of a bcc lattice, with $\vec{q}= (4\pi/a,0,0)$ along with equivalent wave vectors. Of course with NNN hopping the CDW instability
would likely become incommensurate, and would become suppressed as well. Moreover, t}his is not an issue with the fcc lattice, 
as it is not bipartite, the singularity
in the density of states in this instance occurs near the top of the band, and other finite $q$ instabilities will not play such a significant role.
\red{The possibility still remains that a competing $q=0$ instability will suppress superconductivity, but consideration of these is beyond
the scope of this paper.}

Nonetheless, for both the bcc and fcc lattice structures, some NNN hopping is expected, and we have shown here that this immediately
leads to the disappearance of the singularity in the density of states. In fact a `robust' peak remains in the density of states,
and as we have shown, a very large enhancement of $T_c$ continues to be present, now in a regime where the BCS calculation is
more trustworthy, at least for the bcc lattice structure. This is true for several reasons --- for example, as discussed competing
instabilities will be suppressed, but in addition, narrow structures in the density of states will be smeared both by impurities, and
by retardation effects not accounted for in our BCS calculations. Given the number of superconductors with the fcc and bcc lattice structures,
it would be interesting to perform a survey to see if there is any correlation between their critical temperatures and the `remnants' of
these van Hove singularities. In fact, the recent discovery of superconducting hydrogen sulfide under high pressure by
Drozdov et al.\cite{drozdov15} has motivated theoretical work\cite{quan16} that has identified a van Hove singularity in the electronic
density of states. The underlying lattice structure is bcc and thus far the origin of this singularity is not clear. \red{This work, particularly the
previous subsection, strongly suggests that ultimately the origin of the singularity (and ultimately an important factor in the high
superconducting critical temperature) may in fact be BCC structure of the material, along with circumstances that place the Fermi
energy in the vicinity of the robust peak that remains even when NNN hopping is included.}

We have also computed the isotope coefficient in a variety of cases. It is clear that anomalies in this coefficient will exist due
to peaks in the density of states. In some respects the few observations where anomalies are found in known superconductors
can be regarded as signatures of peaks in the density of states (although of course other explanations also exist). We have also 
briefly examined the pairing gap, but there is no significant deviation from what standard BCS theory predicts, i.e. the gap tracks 
the critical temperature.

\begin{acknowledgments}

This work was supported in part by the Natural Sciences and Engineering
Research Council of Canada (NSERC). TXRS is a recipient of an "Emerging Leaders in the Americas Program" (ELAP) scholarship 
from the Canadian government, and we are grateful for this support.

\end{acknowledgments}

\appendix

\section{Density of States within Tight-Binding}

The general equation for the Density of States (DOS) is
\be
g(\epsilon) = {1 \over N} \sum_{k \in {\rm FBZ}} \delta(\epsilon - \epsilon_k),
\label{dos}
\ee
where $\epsilon_k$ is the dispersion relation, the summation is over all points in the First Brillouin Zone (FBZ), and $N$ is the number of $k$-points in the
FBZ. Dispersion relations are determined by overlap integrals and geometry; for a Bravais lattice the dispersion relation can be written as
\be
\epsilon_k = -\sum_\delta t_\delta {\rm cos} \vec{k} \cdot \vec{\delta}
\label{disp}
\ee
where the sum is over all neighbours of a particular lattice site, with decreasing amplitude $t_\delta$, to reflect the decreasing overlap between atoms
that are further apart from one another. This decrease with distance is usually exponential, so very often only nearest neighbour overlaps are considered
to be non-zero. It turns out that this simplified model often possesses special symmetries, not necessarily inherent in the more general model, and therefore,
if for no other reason, further than nearest neighbour overlaps are often considered as well.
Within the tight binding approach only a few nearest neighbours are retained, so for example, 
we obtain
\bea
\epsilon_k &=& -2t\left[ {\rm cos}(k_xa) + {\rm cos}(k_ya) \right] \nonumber \\
 && -4t_2{\rm cos}(k_xa) {\rm cos}(k_ya)  \phantom{aaaaaa} {\rm [2D \ \ NNN]} \\
\epsilon_k &=&  -2t_{s}\left[ {\rm cos}(k_xa) + {\rm cos}(k_ya) +{\rm cos}(k_za) \right]  \nonumber \\ 
&& -4t_{s2} \biggl[ {\rm cos}({k_xa}) {\rm cos}({k_ya}) + {\rm cos}({k_xa}) {\rm cos}({k_za})  \nonumber \\
&& \phantom{aaa} + {\rm cos}({k_ya}) {\rm cos}({k_za})  \biggr] \phantom{aaaaaaa} {\rm [sc \ \ NNN]} \\
\epsilon_k &=& -8t_b\left[ {\rm cos}({k_xa \over 2})  {\rm cos}({k_ya \over 2}) {\rm cos}({k_za \over 2}) \right] \phantom{aa} {\rm [bcc \ \ NNN]} \nonumber \\
& & -2t_{b2}\left[ {\rm cos}(k_xa) + {\rm cos}(k_ya) +{\rm cos}(k_za) \right] \\
\epsilon_k &=& -4t_f \biggl[ {\rm cos}({k_xa \over 2}) {\rm cos}({k_ya \over 2}) + {\rm cos}({k_xa \over 2}) {\rm cos}({k_za \over 2})  \nonumber \\
&& \phantom{aaa} + {\rm cos}({k_ya \over 2}) {\rm cos}({k_za \over 2})  \biggr] \phantom{aaaaaaaaaaa} {\rm [fcc \ \ NNN]} \nonumber \\
& & -2t_{f2}\left[ {\rm cos}(k_xa) + {\rm cos}(k_ya) +{\rm cos}(k_za) \right].
\label{app_examples}
\eea
We repeat here important definitions already mentioned in the text. The distance $a$ is the nearest neighbour distance in the 2D and simple cubic (sc) cases, 
and is the length of the cube in the body-centred cubic (bcc) and face-centred cubic (fcc) cases; these latter two each 
contain 8 atoms, one at each vertex, along with one in the centre (bcc) and six on the face centres (fcc). Also, $t, t_s, t_b$, and $t_f$ are the nearest neighbour hopping parameters and $t_2$, $t_{s2}$, $t_{b2}$, and $t_{f2}$ are the next-nearest neighbour hopping parameters for the 2D square, 3D sc,
3D bcc, and 3D fcc lattices, respectively. 

Note that without NNN hopping these have bandwidths $W$ of $8t$, $12t_s$, $16t_b$, and $16t_f$, respectively. 
We have additionally included next-nearest neighbour hopping in all these cases; for the 3D cases all but the sc case exhibit singularities
when only nearest-neighbour hops are considered; in two dimensions the existence of a singularity is retained as next-nearest neighbour hops are introduced, while in the three dimensions, in either the face-centred cubic (fcc) or body-centred cubic (bcc) cases, the singularity disappears.

For our purposes the important property emerging from these different band dispersions is the shape of the density of states, defined above. In two dimensions, the
DOS can be determined analytically in terms of complete elliptic integrals. The result is 
\be
g_{\rm 2D}(\rho; \epsilon) = {1 \over 2 \pi^2 t a^2} {1 \over \sqrt{1 - 4 \rho \bar{\epsilon}}} K\left[ 1 - {\left(\rho - \bar{\epsilon}\right)^2 \over 1 - 4 \rho \bar{\epsilon}}\right],
\label{dense_2d_nnn}
\ee
where $\bar{\epsilon} \equiv \epsilon/(4t)$ and $\rho \equiv t_2/t$, with the restriction that $-1/2 < \rho < 1/2$. A different expression applies for $|\rho| > 1/2$, but
\begin{figure}[tp]
\begin{center}
\includegraphics[height=3.4in,width=3.0in,angle=-90]{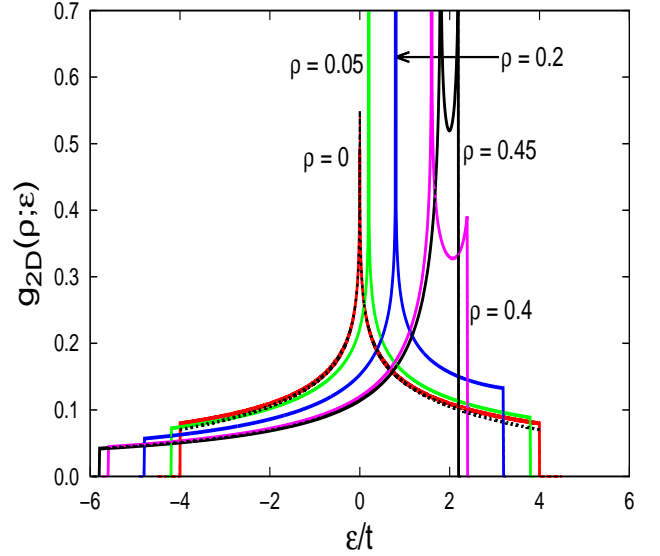}
\end{center}
\caption{Plot of the tight-binding 2D density of states for various values of the next-nearest neighbour (NNN) hopping parameter, $\rho
\equiv t_2/t$, as given analytically in Eq.~(\ref{dense_2d_nnn}). A logarithmic singularity remains even in the 
presence of NNN. Note that the results for negative values of $\rho$ are symmetric (about $\epsilon = 0$) to those 
shown with positive values of $\rho$. As mentioned in the text, numerical results, using Eq.~(\ref{num_dense}), are also shown,
and are indistinguishable from the analytical results. The black dashed curve is the approximation given by 
Eq.~(\ref{dense_2d_app}) in the text, valid for $\rho = 0$.}
\label{figA1}
\end{figure}
we omit this regime as being unphysical. Fig.~(\ref{figA1}) shows the DOS for a few values of $\rho > 0$; note that $g_{\rm 2D}(\rho; \epsilon) = g_{\rm 2D}(-\rho; -\epsilon)$, so the DOS with negative values of $\rho$ are mirror images of those shown. It is evident
from Eq.~(\ref{dense_2d_nnn}) that the logarithmic singularity occurs at $\epsilon = 4t_2$.

\begin{figure}[tp]
\begin{center}
\includegraphics[height=3.4in,width=3.0in,angle=-90]{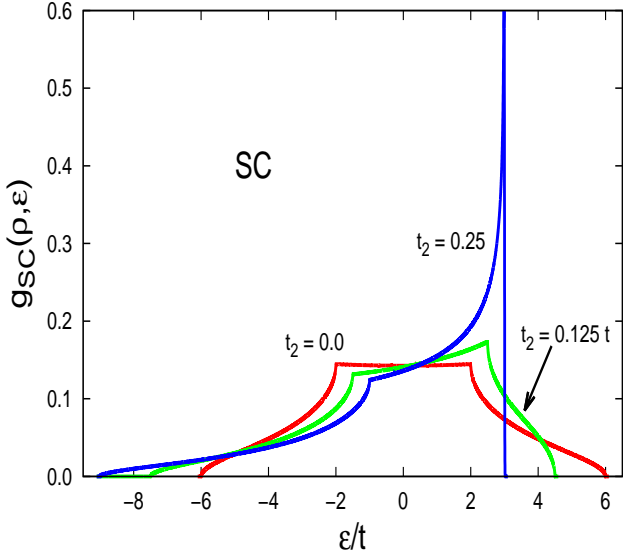}
\end{center}
\caption{Plot of the tight-binding 3D SC density of states for various values of the next-nearest neighbour (NNN) hopping parameter, $\rho
\equiv t_2/t$. Note that a singularity develops at the top of the band, for $\epsilon = 3t$, as the two van Hove singularities (originally at
$\epsilon = 2t$ and at $\epsilon = 6t$ for $t_2 = 0$) merge into one. Results are shown for positive $t_2$ since
the results for negative values of $\rho$ are symmetric (about $\epsilon = 0$) to those 
shown with positive values of $\rho$. }
\label{figAsc}
\end{figure}

\begin{figure}[tp]
\begin{center}
\includegraphics[height=4.0in,width=3.4in,angle=-90]{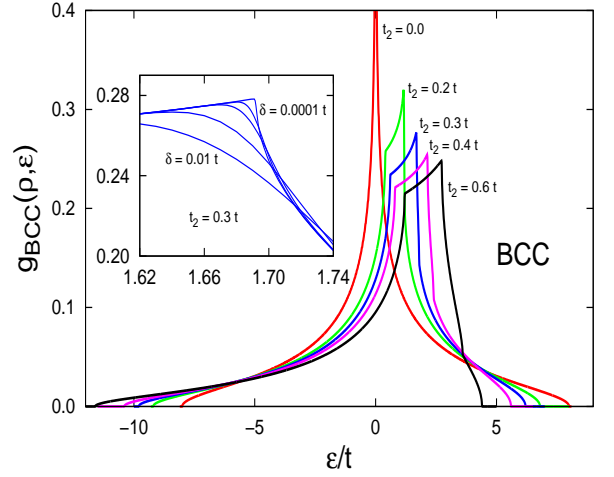}
\end{center}
\caption{Plot of the tight-binding 3D BCC density of states for various values of the next-nearest neighbour (NNN) hopping parameter, $\rho
\equiv t_2/t$. Note that the singularity for $\epsilon = 0$ disappears as $t_2$ becomes non-zero. Results are shown for positive $t_2$ since
the results for negative values of $\rho$ are symmetric (about $\epsilon = 0$) to those 
shown with positive values of $\rho$. Even with non-zero $t_2$ a significant peak in the density of states remains. The inset shows numerical
results as a function of $\epsilon$ very close to the cusp located at $\epsilon_{\rm cusp} = 6t\rho - 4\rho^3t$ for $\rho = t_2/t = 0.3$, for various values of the smearing parameter, $\delta/t = 0.01, 0.005, 0.002, 0.001, 0.0001$.}
\label{figA2}
\end{figure}
\begin{figure}[tp]
\begin{center}
\includegraphics[height=3.2in,width=2.7in,angle=-90]{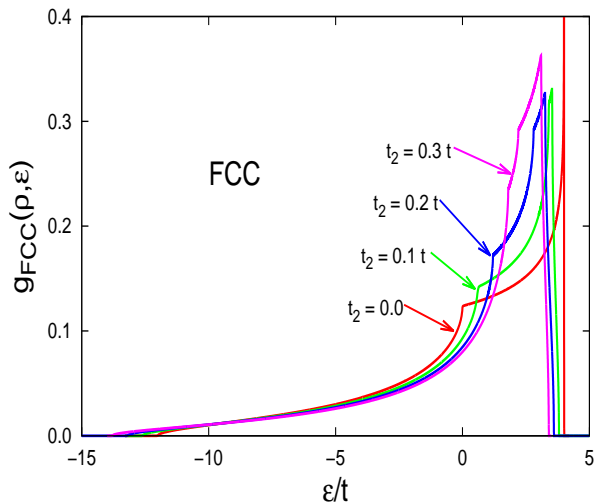}
\end{center}
\caption{Plot of the tight-binding 3D FCC density of states for various values of the next-nearest neighbour (NNN) hopping parameter, $\rho
\equiv t_2/t$. For $t_2 = 0$ there is a singularity at the top of the band (red curve as indicated). As $t_2$ becomes nonzero the singularity disappears and
the maximum shifts to the left. In fact, as $t_2$ grows the maximum in the density of states becomes `robust' in the sense that a significant
area exists in the maximum region (also the case with BCC and with the 2D result --- see the $\rho = 0.45$ result in Fig.~\ref{figA1}).}
\label{figA3}
\end{figure}

Also shown, but indistinguishable from the analytical curves drawn using Eq.~(\ref{dense_2d_nnn}), are results obtained numerically,
using the Gaussian representation for a $\delta$-function. With $x \equiv k_x a/\pi$ and $y \equiv k_y a/\pi$, the 2D version for the
first dispersion given in Eq.~(\ref{app_examples}) is
\be
g_\delta(\epsilon) = {1 \over 2 t a^2} {1 \over \sqrt{\pi \delta^2}} \int_0^1  dx \int_0^1 dy \ {\rm exp}
\left\{-\left[{\epsilon - \epsilon_k \over 2t\delta}\right]^2\right\},
\label{num_dense}
\ee
where the approximation improves for smaller value of the smearing parameter, $\delta$. In Fig.~(\ref{figA1}) we use
$\delta = 0.0005t$. For 3D dispersions an additional integral over $z \equiv k_z a/\pi$ from zero to unity is required and $a^2 \rightarrow a^3$.

In three dimensions, the integrals must be done numerically. It is straightforward to simplify some of the results when there is no
next-nearest-neighbour hopping, \cite{jelitto69} and we cite some of these results for convenience. For the rest we develop formulas in some
instances or simply use Eq.~(\ref{num_dense}),
as this is straightforward and continues to work extremely well, even in 3 dimensions.

The result for SC with next-nearest-neighbour hopping is
\be
g_{\rm SC}(\epsilon) = \int {dx \over 2 \pi^2 ta^3} {K(y) \over \sqrt{(1 + 2 \rho {\rm cos} \pi x)^2 - \rho(4 \bar{\epsilon} + 2 {\rm cos} \pi x)}},
\label{sc_anal}
\ee
where
\be
y \equiv 1-{(\rho - \bar{\epsilon} -{1 \over 2}{\rm cos} \pi x)^2 \over (1 + 2 \rho {\rm cos} \pi x)^2 - \rho(4 \bar{\epsilon} + 2 {\rm cos} \pi x)}
\label{yy}
\ee
and $\bar{\epsilon} \equiv \epsilon/(4t)$, $K(y)$ is the complete elliptic integral of the first kind, and $\rho \equiv t_2/t$ is the ratio of the NNN to
the NN hopping amplitude, and $a^3$ is the unit cell volume. The limits on the integration are such that $y$ remains positive and less than unity
at all times. As remarked in the text, this density of states has no singularities for $t_2 = 0$ (there remain van Hove singularities in the form of cusps
and singularities in the derivative of $g(\epsilon)$), but develops a singularity as $t_2$ increases. This is evident in Fig.~\ref{figAsc}, where we use
the 3D version of Eq.~(\ref{num_dense}) to plot the density of states.

The result for the BCC lattice with nearest-neighbour hopping only is\cite{jelitto69} 
\be
g_{\rm BCC}(\epsilon) = {2 \over a^3} {1 \over 2 \pi^3 t}\int_{|\bar{\epsilon}|}^1 dx \ {1 \over \sqrt{x^2 - \bar{\epsilon}^2}} K\left[1-x^2\right].
\label{bcc_anal}
\ee
Note that the unit cell volume for the BCC lattice is $a^3/2$; that is why we isolate this factor at the front of the previous formula.
Using the fact that $K(1-x^2) \rightarrow {\rm ln}(4/x)$ as $x \rightarrow 0$, one can straightforwardly derive that
\be
\lim_{\bar{\epsilon} \rightarrow 0} g_{\rm BCC}(\epsilon) = {2 \over a^3}{1 \over 2\pi^3 t} \left[ {3 \over 2} {\ln}^2({1 \over |\bar{\epsilon}|})
+ 3\ {\rm ln} 2 \ {\rm ln}({1 \over |\bar{\epsilon}|}) + 2({\rm ln}2)^2 \right],
\label{asym}
\ee
so that the divergence is a $\approx {\rm ln}^2\left({1\over |\bar{\epsilon}|}\right)$, singularity, stronger than occurs in two dimensions. For the case with NNN hopping
we use Eq.~(\ref{num_dense}) to determine the result numerically; these are shown in Fig.~(\ref{figA2}). The van Hove points are at energies $-8t -6t\rho$
(bottom of the band), $2t\rho$, $6t\rho - 4t\rho^3$, $6t\rho$, and $8t - 6t\rho$ (top of the band). Note that the bandwidth remains $16t$ even when NNN hopping is non-zero.

Finally, for FCC, we show numerical results for NNN hopping as well. Our use of the exponential representation of the $\delta$-function still allows
sufficient resolution to show the various van Hove singularities, apparent in Fig.~(\ref{figA3}). However, with NNN hopping the singularity at the top of the
band for $t_2=0$ disappears, but various cusps remain, signifying discontinuities in the first derivative. Note that the FCC lattice is not bipartite, and 
nesting is not present, even in the case of NN hopping only.


\begin{thebibliography}{99}

\bibitem{bardeen57} J. Bardeen, L.N. Cooper and J.R. Schrieffer, {\it Theory of Superconductivity},
Phys. Rev. {\bf 106}, 162 (1957); Phys. Rev. {\bf 108}, 1175 (1957).

\bibitem{labbe67} J. Labb\'e. S. Bari\v si\'c and J. Friedel, {\it Strong-coupling Superconductivity in V$_3$X Type of Compounds}, 
Phys. Rev. Lett. {\bf 19}, 1039 (1967).

\bibitem{nettel77} S.J. Nettel and H. Thomas, {\it Electron-density of States and Superconducting $T_c$ in A15-Compounds},
Solid State Commun. {\bf 21} 683 (1977).

\bibitem{horsch77} P. Horsch and H. Rietschel, {\it New Aspect of Superconductivity in A-15 Compounds}, 
Z. Phys. B {\bf 27} 153 (1977).

\bibitem{lie78} S.G. Lie and J.P. Carbotte, {\it Dependence of $T_c$ on Electronic Density of States}. 
Solid State Commun. {\bf 26} 511 (1978).

\bibitem{ho78} K.M. Ho, M.L. Cohen, and W.E. Pickett, {\it Maximum Superconducting Transition-Temperatures in A15 Compounds},
Phys. Rev. Lett. {\bf 41} 815 (1978).

\bibitem{pickett80} W.E. Pickett, {\it Effect of a Varying Density of States on Superconductivity},
Phys. Rev. B {\bf 21} 3897 (1980).

\bibitem{mitrovic83} B. Mitrovi\'c and J.P. Carbotte, {\it Effects of Energy-Dependence in the Electronic Density of States on some Normal State
Properties}, Can. J. Phys. {\bf 61} 758 (1983); {\it Effects of Energy-Dependence in the Electronic Density of States on some Superconducting
Properties},  Can. J. Phys. {\bf 61} 784 (1983); {\it Free-Energy Formula for a Strong Coupling Superconductor with Energy-dependent Electronic Density of
States}, Can. J. Phys. {\bf 61} 872 (1983).

\bibitem{stewart15} See, for example, the very recent review by G.R. Stewart, Physica C {\bf 514}, 28-35 (2015), titled "{\it Superconductivity in the A15
Structure}". This is a review in the Special Issue on Superconducting Materials, edited by J.E. Hirsch, M.B. Maple and F. Marsiglio.

\bibitem{hirsch86} J.E. Hirsch and D.J. Scalapino, {\it Enhanced Superconductivity in Quasi Two-dimensional Systems},
Phys. Rev. Lett. {\bf 56}, 2732 (1986).

\bibitem{labbe87} J. Labb\'e and J. Bok, \com{\itt Superconductivity in Alcaline-Earth-Substituted La$_2$CuO$_4$: a Theoretical Model,} 
Europhys. Lett. {\bf 3}, 1225-1230 (1987).

\bibitem{tsuei90} C.C. Tsuei, D.M. Newns, C.C. Chi and P.C. Pattnaik, \com{\itt Anomalous Isotope Effect and van Hove Singularity in Superconducting Cu Oxides,} 
Phys. Rev. Lett. {\bf 65}, 2724-2727 (1990).

\bibitem{markiewicz97} R.S. Markiewicz, \com{\itt A Survey of the van Hove Scenario for High-T$_c$ Superconductivity with Special
Emphasis on Pseudogaps and Striped Phases,} J. Phys. Chem. Solids {\bf 58}, 1179-1310 (1997).

\bibitem{bok12} J. Bok and J. Bouvier, \com{\itt Superconductivity and the van Hove Scenario,} J. Supercond. Nov. Magn. {\bf 25}, 657-667 (2012).

\bibitem{andersen91} O.K. Andersen, A.I. Liechtenstein, O. Rodriguez, I.I. Mazin, O. Jepsen, V.P. Antropov, O. Gunnarsson, and S. Gopalan,
\com{\itt Electrons, phonons, and their interaction in YBa$_2$Cu$_3$O$_7$,} Physica C {\bf 185-189}, 147-155 (1991).

\bibitem{abrikosov93} A.A. Abrikosov, J.C. Campuzano, and K. Gofron, \com{\itt Experimentally observed extended saddle point singularity in the energy spectrum of YBa2Cu3O6.9 and YBa2Cu4O8 and some of the consequences,} Physica C {\bf 214}, 73-79 (1993).

\bibitem{jelitto69} R.J. Jelitto, \com{\itt The Density of States of some Simple Excitations in Solids,} J. Phys. Chem. Solids {\bf 30}, 609-626 (1969).

\bibitem{duan14} \red{Defang Duan, Yunxian Liu, Fubo Tian, Da Li, Xiaoli Huang, Zhonglong Zhao, Hongyu Yu, Bingbing Liu, Wenjing Tian, and Tian Cui,
\com{\itt Pressure-induced metallization of dense (H$_2$S)$_2$H$_2$ with high-T$_c$ superconductivity,} Scientific Reports {\bf 4}, 6968-1-6 (2014).}

\bibitem{bernstein15} \red{N. Bernstein, C. Stephen Hellberg, M.D. Johannes, I.I. Mazin, and M.J. Mehl, \com{\itt What superconducts in sulfur hydrides
under pressure and why,} Phys. Rev. B{\bf 91}, 060511(R)-1-5 (2015).}

\bibitem{drozdov15} A.P. Drozdov,  M.I. Eremets, I.A. Troyan, V. Ksenofontov, and S.I. Shylin,
{\it Conventional superconductivity at 203 kelvin at high pressures in the sulfur hydride system}, Nature {\bf 525}, 73-76 (2015).

\bibitem{remark0} \red{We became aware of Ref. [\onlinecite{sano16}] only after this paper was initially submitted; they have 
investigated retardation effects and indeed find that a reduction in T$_c$ will occur as a result of their inclusion.}

\bibitem{nozieres85} P. Nozi\`eres and S. Schmitt--Rink, {\it Bose Condensation in an Attractive Fermion Gas - From Weak to Strong Coupling
Superconductivity}, J. Low Temp. Phys. {\bf 59}, 195 (1985).

\bibitem{marsiglio15} A recent description of the two-dimensional electron gas with attractive interactions within the T-matrix formalism is
given in F. Marsiglio, P. Pieri, A. Perali, F. Palestini, and G.C. Strinati,  \com{\itt Pairing effects in the normal phase of a two-dimensional Fermi gas,}
Phys. Rev. B{\bf 91}, 054509 (2015).

\bibitem{marsiglio92} F. Marsiglio, \com{\itt Eliashberg Theory of the Critical Temperature and Isotope Effect. Dependence on Bandwidth,
Band-Filling, and Direct Coulomb Repulsion,} J. Low Temp. Phys. {\bf 87} 659-682 (1992).

\bibitem{marsiglio08} F. Marsiglio and J.P. Carbotte, `Electron-Phonon Superconductivity',
Review Chapter in {\it Superconductivity, Conventional and Unconventional Superconductors},
edited by K.H. Bennemann and J.B. Ketterson (Springer-Verlag, Berlin, 2008), pp. 73-162.


\bibitem{eliashberg60} G.M. Eliashberg, \com{\itt Interactions between Elecrons and Lattice Vibrations in a Superconductor,} Zh. Eksperim. i Teor. Fiz.
{\bf 38} 966 (1960); Soviet Phys.  JETP {\bf 11} 696-702 (1960).

\bibitem{quan16} Y. Quan and W.E. Pickett, {\it Van Hove singularities and spectral smearing in high-temperature superconducting H$_3$S},
Phys. Rev. B{\bf 93}, 104526 (2016).

\bibitem{sano16} \red{Wataru Sano, Takashi Koretsune, Terumasa Tadano, Ryosuke Akashi, and Ryotaro Arita, {\it Effect of Van Hove singularities
on high-T$_c$ superconductivity in H$_3$S}, Phys. Rev. B{\bf 93}, 094525-1-16 (2016).}

\bibitem{schrieffer64} J.R. Schrieffer, In: {\it Theory of Superconductivity}
(Benjamin/Cummings, Don Mills, 1964).

\bibitem{tinkham96} M. Tinkham, In: {\it Introduction to Superconductivity},
(Second Edition, McGraw--Hill, New York, 1996).

\bibitem{remark1} In this manner retardation effects are taken into account in BCS theory in a very phenomenological
way through the imposed cutoff. This cutoff is imposed in momentum space (not in frequency space); nonetheless, use
of a simple identity for the factor $1-2f(E_k)$ allows the right-hand-side of the so-called gap equation [Eq. (\ref{bcs1})]
to be rewritten in terms of Matsubara frequencies with a cutoff in this space, so it better resembles Eliashberg theory.

\bibitem{nist10} Frank W.J. Olver, Daniel W. Lozier, Ronald F. Boisvert, and Charles W. Clark, {\it NIST Handbook
of Mathematical Functions}, (Cambridge University Press, Cambridge, 2010).

\bibitem{remark2} \red{One of us (FM) has repeatedly asked colleagues in the superconducting community
around the world about this for the last 20 years, and with one exception they were not aware of the divergences in three dimensions.
The Jelitto paper\cite{jelitto69} has been cited more than 140 times, so clearly some researchers are aware of this fact. However, neither
of Refs. [\onlinecite{duan14,bernstein15}] cite the Jelitto paper. While later references\cite{papa15,quan16} cite a van Hove singularity, neither
of these seem to be aware that the relevant van Hove singularity is likely a remnant of the BCC singularity (when only NN hopping is included).
Ref. [\onlinecite{ortenzi16}] actually uses a rather sophisticated tight-binding model to achieve a good fit with first-principle calculations, but they
also appear to be unaware of the BCC singularity lurking nearby in parameter space.}

\bibitem{papa15} \red{D.A. Papaconstantopoulos, B.M. Klein, M.J. Mehl, and W.E. Pickett, {\it Cubic H$_3$S around $200$ GPa: An 
atomic hydrogen superconductor stabilized by sulfur,} Phys. Rev. B{\bf 91}, 184511-1-5 (2015).}

\bibitem{ortenzi16} \red{Luciano Ortenzi, Emmanuele Cappelluti, and Luciano Pietronero, {\it Band structure and electron-phonon coupling 
in H$_3$S: a tight-binding model}, arXiv:1511.04304.}

\bibitem{marsiglio90a} F. Marsiglio, {\it Pairing and charge-density-wave correlations in the Holstein model at half-filling},
Phys. Rev. B{\bf 42}, 2416-2424 (1990).

\bibitem{kosterlitz73} J.M. Kosterlitz and D.J. Thouless, {\it Ordering, Metastability and Phase-transitions in 2-Dimensional Systems}, 
J. Phys. {\bf C6}, 1181 (1973).




%
%
%
%
%
%
%
%
%
%
%
%
%
%
%
%
%
%
%
%
%
%
%
%
%
%













\end{thebibliography}
\end{document}